\documentclass[fleqn,usenatbib]{mnras}

\usepackage{natbib} 
\bibpunct{(}{)}{;}{a}{}{,} 
\usepackage{graphicx}
\usepackage{txfonts}
\def\inte{{\em INTEGRAL}}

\def\suzaku{{\em Suzaku}}

\def\porb{P_{\rm orb}} 
 
\def\mstar{M_{\rm *}}
\def\mns{M_{\rm NS}}
\def\vx{v_{\rm x}}
\def\vy{v_{\rm y}}
\def\vw{v_{\rm w}}
\def\rstar{R_{\rm *}}
\def\lstar{L_{\rm *}}
\def\lsun{L_{\odot}}
\def\msun{M_{\odot}}
\def\dotm{\dot{M}}
\def\rsun{R_{\odot}}

   \title[Wind accretion in neutron star supergiant high mass X-ray binaries]{A semi-analytical treatment to wind accretion in neutron star supergiant high mass X-ray binaries: I. eccentric orbits}

\author[E. Bozzo et al.]{
E. Bozzo,$^{1}$\thanks{E-mail: enrico.bozzo@unige.ch}
L. Ducci,$^{1,2}$
and M. Falanga$^{3}$
\\
$^{1}$Department of Astronomy, University of Geneva, Chemin d'Ecogia 16, CH-1290 Versoix, Switzerland\\
$^{2}$Institut f\"ur Astronomie und Astrophysik, Kepler Center for Astro and Particle Physics, Eberhard Karls Universit\"at, Sand 1, 72076 T\"ubingen, Germany\\
$^{3}$International Space Science Institute (ISSI) Hallerstrasse 6, CH-3012 Bern, Switzerland
}

\date{}
   
\begin{document}
\label{firstpage}
\pagerange{\pageref{firstpage}--\pageref{lastpage}}
\maketitle

\begin{abstract}
We present in this paper a first step toward a semi-analytical treatment of the accretion process in wind-fed neutron star supergiant X-ray binaries with eccentric orbits. We consider the case of a spherically symmetric wind for the supergiant star and a simplified model for the accretion onto the compact object. A self-consistent calculation of the photoionization of the stellar wind by the X-rays from the accreting neutron star is included. This effect is convolved with the modulation of the mass accretion rate induced by the eccentric orbit to obtain the expected X-ray luminosity of a system along the orbit. As part of our results, we first show that the bi-modality of low and high X-ray luminosity solutions for supergiant X-ray binaries reported in previous papers is likely to result from the effect of the neutron star approaching first and then moving away from the companion (without coexisting simultaneously). We propose that episodes of strong wind photoionization can give rise to off-states of the sources. Our calculations are applied to the case of a few classical supergiant X-ray binary systems with known eccentricities (Vela\,X-1, 4U\,1907+09, GX\,301-2) and to the case of the only supergiant fast X-ray transient with a confirmed eccentric orbit, IGR\,J08408-4503. The results are compared with observational findings on these sources. We also discuss the next steps needed to expand the calculations toward a more comprehensive treatment in future publications.
\end{abstract}   
  

\section{Introduction}
\label{sec:intro} 

High mass X-ray binaries (HMXBs) comprise a compact object, usually a neutron star (NS), that accretes matter from a massive companion ($\mstar$$\gg$1$\msun$). The so-called supergiant high mass X-ray binaries (SgXBs) are a sub-class of the HMXBs in which the companion star is an OB supergiant and accretion occurs preferentially through the stellar wind \citep{walter2015}. 

HMXBs and SgXBs have been largely investigated in the past decades as these are among the brightest X-ray sources in our Galaxy \citep[see, e.g.,][for an historical review]{heuvel91}. SgXBs have gained a remarkably renewed interest in the most recent years due to the fact that, beside being unique laboratories to study the physics of accretion and the condition of matters under extreme magnetic and gravitational fields, the NSs in these systems provide a crucial tool to probe \emph{in situ} the physical properties of supergiant stellar winds. The basic properties of radiatively accelerated winds from OB supergiants are relatively well understood nowadays, but there are many fundamental aspects that still need to be clarified.  Discrepancies between theoretical studies (including multi-dimensional hydrodynamic simulations) and observations in both the UV and X-ray domain have been reported \citep{nunez17}. 

The X-ray emission released by a NS that is accreting from the wind of a supergiant is readily affected by changes in the velocity and density of the wind. The passage through regions with different material densities and velocities results in an immediate variation (within few to thousands of seconds) of the released energy.  The induced X-ray variability can thus be used, in principle, as a tool to evaluate the physical conditions of the stellar wind matter that is crossed by the NS. Although the method has been proven promising, \citet{nunez17} discussed that a quantitative estimate of the stellar wind properties from the X-ray observations of SgXBs is currently hampered by the impossibility of taking into account in the calculations all uncertainties associated with the accretion process. The luminosity released during the wind accretion onto the NS cannot be trivially associated to density/velocity variations in the stellar wind as this luminosity is the result of a complex interplay between numerous physical processes that can significantly boost or halt the X-ray variability \citep[see discussion in][and references therein]{bozzo15}. Among the most relevant processes there is the photoionization of the stellar wind by the NS X-rays which can drastically slow down the wind, which in turns affects the released high energy radiation from the compact object. A major role is also played by the plasma-magnetic field interactions occurring at the NS magnetospheric boundary, where the dynamics of the inflowing stellar wind matter is controlled by the NS intense magnetic field and rotation \citep[see, e.g.,][and references therein]{bozzo08, shakura12}. 

One of the first semi-analytical calculations about the stellar wind photoionization in HMXBs was presented by \citet{ho87} who found a way to construct a feedback equation taking into account the released X-ray luminosity and the corresponding influence on the stellar wind acceleration. These authors suggested a bi-modality in the accretion states of wind-fed HMXBs, showing that there are two branches of solutions to the feedback equation: one resulting in a slow wind and high X-ray luminosity, and one corresponding to the case of a fast wind and low X-ray luminosity. Pioneering numerical calculations on this aspect were presented later by \citet{stevens90} and \citet{stevens91} using parametric wind models. These studies were more recently evolved by \citet{krtika12} who introduced the usage of 2D wind models adopting a calculation method in which the NS is placed at a certain distance from the massive companion neglecting the orbital revolution. They confirmed the presence of the two branches of solutions originally discussed by \citet{ho87} and provided in a later paper a full exploration of the changes in the photoionization effect for different distances between the massive companion and the compact object as a function of the X-ray luminosity \citep{krtika15,krtika16}. \citet{karino14} performed a further step forward by introducing in a similar calculation as that presented by \citet{krtika12} the possibility of estimating the X-ray luminosity self-consistently within the model rather than imposing this as an input parameter. This author also did not take into account the orbital motion of the NS and confirmed again the presence of the two branches of solutions. It was  suggested  that the high X-ray luminosity solution could be applicable to the case of the so-called persistent (or ``classical'') SgXBs, while the low luminosity solution was to be used to interpret the supergiant fast X-ray transients \citep[SFXTs;][]{negueruela05, sguera06}. These are a sub-class of the SgXBs showing a remarkable variability in the X-rays that is not completely well understood but can achieve factors of 10$^4$ higher than that of classical SgXBs  \citep[see, e.g.,][and references therein]{nunez17}. In all these calculations, the wind of the massive star is always assumed to be smooth and spherically symmetric and only more recently \citet{krtika18} introduced the effect of wind clumping on the photoionization calculations, showing that clumps generally tend to lower the dumping of the wind velocity by the X-ray luminosity compared to the smooth wind case. 

\citet{manousakis12} used 2D hydrodynamic simulations to estimate the luminosity resulting from the accretion of a smooth and symmetric stellar wind onto a NS in a circular orbit around its supergiant companion, providing interesting insights on the X-ray variability induced by instabilities within the hydrodynamic flow. Magnetospheric interactions are not included in these simulations due to the current impossibility of developing a complete magneto-hydrodynamic numerical code for this complex problem. These authors did, however, include a simplified treatment of the stellar wind photonization by the X-ray photons from the NS and the effect of the Coriolis' force. A different simulation approach was undertaken by \citet{mellah18}, who used an evolved 2D hydrodynamic model of stellar winds with reasonably sized clumps and computed 3D simulations of the accretion of these clumps onto a NS immersed in this wind \citep[see also][]{karino15}.  The computation included the effect of wind photoionization in a simplified fashion and that of gravity. Similarly to the case of \citet{manousakis12}, also \citet{mellah18} could not take into account the effect of magnetospheric interactions. Although the progresses on numerical simulations are remarkable \citep[see also][]{mellah20}, they are still assuming simplified scenarios and cannot generally provide scalable formulas that can be used to interpret observational results (mainly in the X-ray domain). 

Calculations of the clumpy wind accretion onto a NS where the magnetospheric interactions are taken into account where presented by \citet{bozzo16}, extending a previous work published by \citet{oskinova2012}. The authors first evaluated the effect of a 1D clumpy stellar wind on the X-ray luminosity released by an accreting NS immersed in this wind and also included the complication of the magnetospheric interactions. They showed that the behavior of classical SgXBs and SFXTs could be reproduced by exploring the space of the model parameters but also highlighted that their approach was simplified in several respects: (i) the 1D clumpy wind model bears known limitations and could overestimate the size of the clumps compared to multi-dimensional approaches \citep[see, e.g.,][]{dessart02, dessart03, dessart05}; (ii) the effect of the feedback from the X-ray radiation onto the stellar wind, as well as (iii) the Coriolis' force resulting from the orbital motion and the NS gravitational pull were not included (they assumed strictly circular orbits). 

As part of this extensive effort toward a better understanding of the wind accretion process and the relevance of different effects impacting on the estimates of the physical properties of supergiant stellar winds from observations of SgXBs, we present here the first attempt to construct a semi-analytical calculation of the interaction between the stellar wind from an OB supergiant with the orbiting NS. We consider the simplified case of a smooth and symmetric wind, and assume a simplified scenario for the magnetospheric interactions. We do however include the effect of eccentric orbits, that was never considered in previous approaches. The unique advantage of the proposed calculation approach is the possibility to progressively include all complications and provide relatively simple equations that can be easily scaled and applied to observed NS SgXBs. Although the semi-analitical approach requires inevitable simplifications at this stage, all assumptions are clearly spelled out in the paper and their effect thoroughly analyzed. We summarize in Sect.~\ref{sec:model} our considered assumptions for the stellar wind and the NS orbit, while in Sect.~\ref{sec:simpler} we present all the equations of our considered simplified wind accretion scenario.  
We discuss the main findings of our calculations in the context of previous literature results and relevant observations of SgXBs in Sect.~\ref{sec:results}. Our conclusions and comments on the next steps forward are reported in Sect.~\ref{sec:conclusions}. We plan to gradually increase the complexity of the calculations presented here in forthcoming publications including first the effect of a clumpy wind and then of detailed magnetospheric interactions. 

\section{Basic model assumptions}
\label{sec:model} 

\begin{figure}
\centering
\includegraphics[scale=0.5]{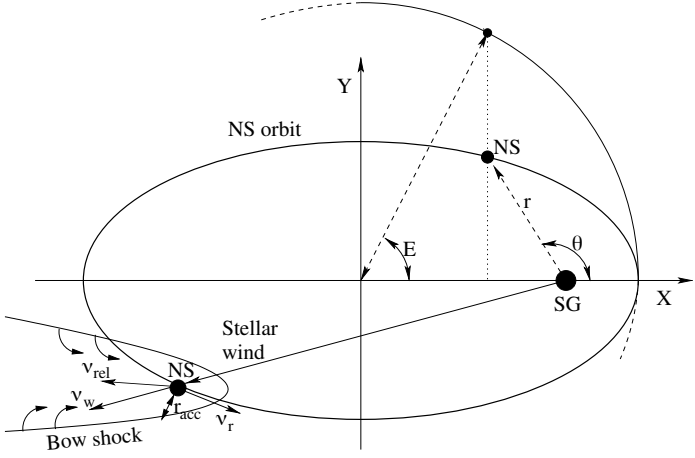}
\caption{A sketch of the NS orbit around the supergiant companion. The most relevant coordinates are indicated together with the eccentric anomaly. C represents the center of the $(x,y)$ coordinate system adopted throughout this paper. The vectors corresponding to the wind, as well as to the orbital and relative velocity between the NS and the supergiant, are also represented. This figure sets the reference frame used for all equations in this paper (unless specified otherwise). The supergiant star is fixed in this reference frame.} 
\label{fig:coord1} 
\end{figure}

In this paper we consider the simplified case of a NS orbiting an OB supergiant star in an eccentric orbit and accreting from its intense wind. The stellar wind is assumed to be smooth and symmetric, according to the CAK prescription \citep{cak1975}. We also adopt a simplified accretion scenario, following the approach presented by, e.g., \citet{davies79}. We also neglect the effect of the Coriolis' force in order to be able to analytically solve most of the equations. Neglecting the Coriolis' force does not allow us to study the formation of accretion wakes and the deformation of the shock front around the compact object \citep[see, e.g.,][and references therein]{manousakis12}, but this does not affect our main conclusions. 
The equations we develop in the next sections are generalized and can be applied to systems with different characteristic parameters. The free parameters of the model are the orbital period, the eccentricity, the mass and radius of the NS and of the supergiant star, the magnetic field and spin period of the NS, the supergiant wind terminal velocity, its $\beta$ parameter derived from the classical CAK theory, and the mass loss rate for the supergiant star.  
In Sect.~\ref{sec:orbit} and \ref{sec:wind} we describe in details the basic assumptions used for the description of the Keplerian orbits and for the stellar wind. 
In Sect.~\ref{sec:simpler} we provide the details of the simplified accretion model to compute the system X-ray luminosity as a function of all free parameters.

\subsection{Keplerian orbits}
\label{sec:orbit} 

Throughout this paper, we consider that a NS, with mass $M_{\rm NS}$ and radius $R_{\rm NS}$, is moving around its supergiant companion (mass $M_{\rm NS}$ and radius $R_{\rm NS}$) in an eccentric Keplerian orbit. Both stars are considered point masses. The system is confined in the x-y plane, and the elliptic orbit is defined by the equation:
\begin{equation}
x^2/a^2+y^2/b^2=1. 
\end{equation}  
Here $a$ and $b$ are the semi major and minor axis of the orbit, respectively. 
The eccentricity is defined by $e$=$\sqrt{1-(b/a)^2}$, and we use the relation
\begin{equation}
a=(\porb/2\pi)^{2/3} [G(\mns+\mstar)]^{1/3},  
\end{equation}  
where $\porb$ is the orbital period of the system \citep[see, e.g.,][for the full derivations of the basic equations used in this section]{book12}. 
We also use the definition of the eccentric anomaly, 
$\cos{E}$=$x$/$a$, to define the relation between 
the radial distance $r$ and the semimajor axis: 
\begin{equation}
r=a(1-a\cos{E})=a-ex=\frac{a(1-e^2)}{1+e\cos{\theta}}. 
\label{eq:erre}
\end{equation}  
The most relevant variables are shown in Fig.~\ref{fig:coord1}, which also sets the reference frame considered for all analytical developments in this paper (unless state otherwise). 
The orbital velocity of the NS can be expressed as  
\begin{eqnarray}
\vx & = & -\sqrt{\mu/p}\sin{\theta} \\ 
\vy & = & \sqrt{\mu/p}(e+\cos{\theta}),
\label{eq:orbitalvelocity} 
\end{eqnarray} 
where $\mu$=$G(\mns+\mstar)$ and $p$=$r(1+e\cos{\theta})$. 
In each point of the NS orbit, we consider for simplicity that the  
velocity, $\vec{v}_{\rm w}$, of the supergiant wind is only along $\vec{r}$, and thus 
\begin{eqnarray}
v_{\rm wx} & = & - \vert\vec{\vw}\vert\cos{\theta} \\
v_{\rm wy} & = & \vert\vec{\vw}\vert\sin{\theta} \mbox{ .}
\label{eq:wind}
\end{eqnarray}
The quantity $\vert\vec{v}_{\rm w}\vert$ is discussed in more detail in Sect.~\ref{sec:wind}. 
From Eqs.~\ref{eq:orbitalvelocity} and \ref{eq:wind}, we can now estimate the relative velocity 
between the NS and the stellar wind, $\vec{v}_{\rm rel}$, in each point of the orbit: 
\begin{equation}
\vert{\vec{v}_{\rm rel}}\vert = \sqrt{\mu/p (e^2+1+2e\cos{\theta})-2\sqrt{\mu/p} e\sin{\theta}\vert\vw\vert+\vert\vw\vert^2}.   
\end{equation}  

\begin{figure*}
\centering
\includegraphics[scale=0.35]{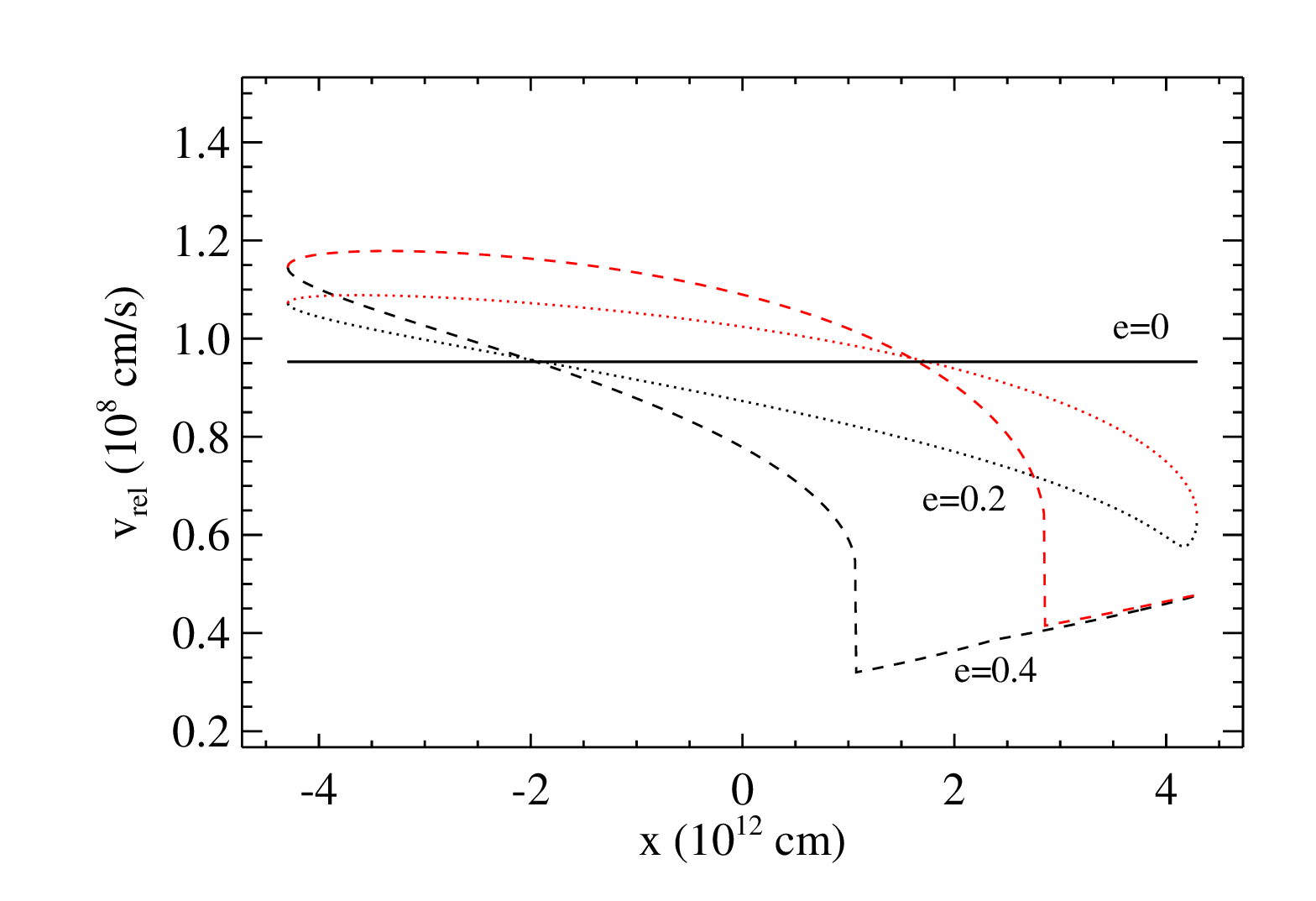}
\includegraphics[scale=0.35]{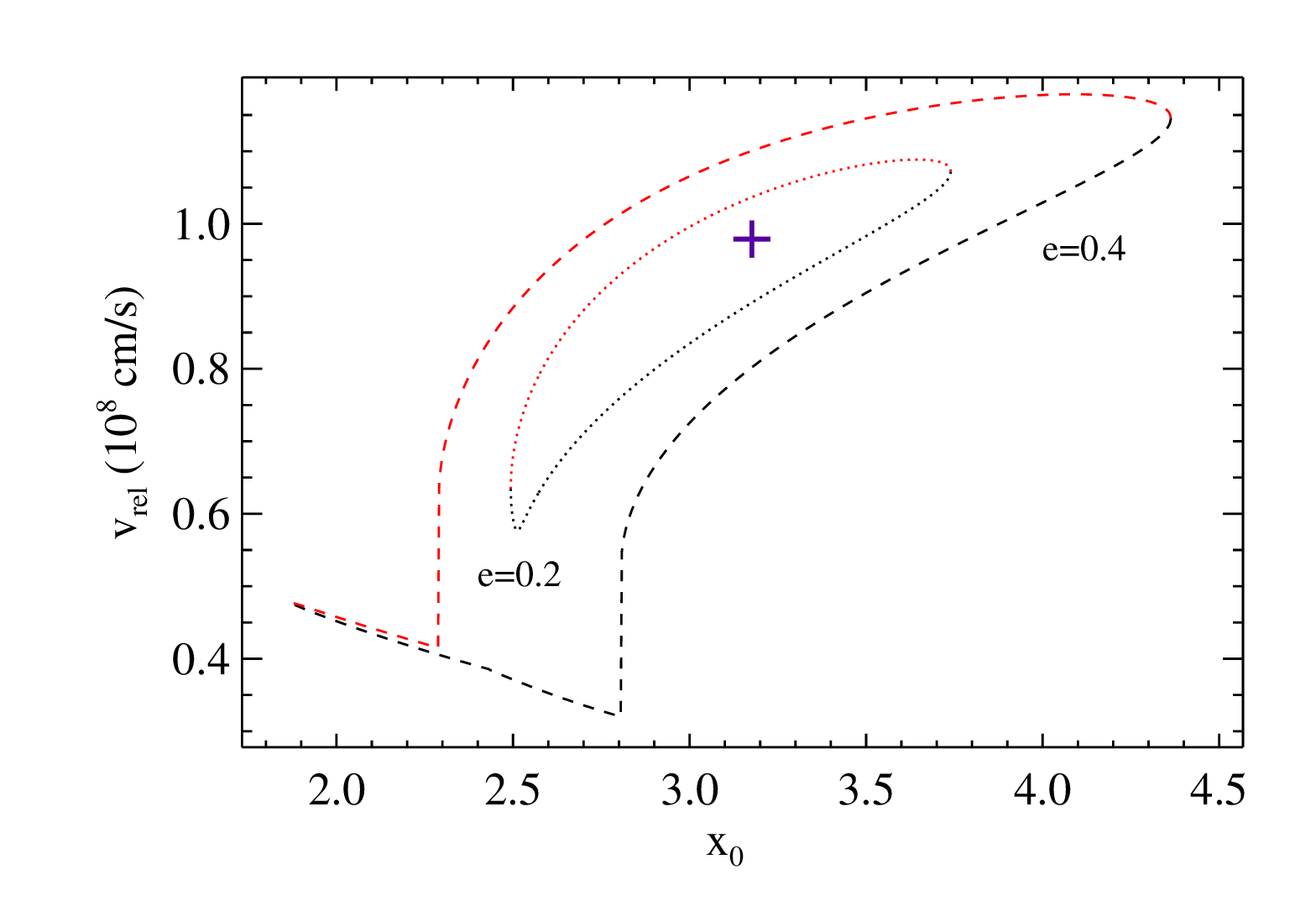}
\caption{{\it Left}: Changes in the relative velocity between the NS and the supergiant wind along the orbit of the compact object. In the plot, $v_{\rm rel}$ is calculated as a function of the $x$ coordinate on the orbital plane. We assumed in this case an orbital period of 10~days, a NS mass (radius) of 1.4\,$M_{\odot}$ (10~km), and parameters for the OB supergiant of $T_{\rm eff}$=33000\,K, $\beta$=1.0, $M_{*}$=30$M_{\odot}$, and $R_{*}$=20$R_{\odot}$ (see Sect.~\ref{sec:wind}). A value of $\xi_{\rm cri}$=100 has been assumed (see Sect.~\ref{sec:ionization2}). The solid line corresponds to the case of a circular orbit ($e$=0), while the dotted (dashed) line represents the case $e$=0.2 ($e$=0.4). {\it Right}: same as the left figure but in this case the relative velocity is plot against the radial distance between the NS and the supergiant companion. The case $e=0$ is represented with a cross, as for a circular orbit the radial separation between the NS and its companion does not change. In both figures on the left and right the anomalous behavior of the relative velocity in the case $e=0.4$ when the NS is close to the supergiant is due to the photoionization of the wind (see Sect.~\ref{sec:ionization2}). Furthermore, we represented in black (red) the branch of values corresponding to the plane $y>0$ ($y<0$) in Fig.~\ref{fig:coord1}, where the NS is moving away (approaching) the supergiant companion.} 
\label{fig:coord} 
\end{figure*}

A typical simplification that is adopted while dealing with with accretion in SgXBs is that $v_{\rm rel}$$\simeq$$v_{\rm w}$, as supergiant stars posses fast winds and the orbital velocity of the NS is small compared to it for orbital periods of several days. However, we show in Fig.~\ref{fig:coord} that the above approximation is no longer applicable when 
a non-negligible eccentricity is introduced. In particular, at higher eccentricities the profile of $v_{\rm rel}$ along the orbit gets more and more asymmetric while the NS approaches and then moves away from the companion. 

\subsection{The supergiant wind}
\label{sec:wind} 

In order to define the functional form of $\vert\vw\vert$, we follow the treatment in \citet{lamers99} and adopted the simplified approach according to which the stellar wind velocity is escaping radially from the surface of the supergiant star (assumed to be spherically symmetric).  Observations and models of winds from supergiant stars indicate that the wind velocity can often be approximated by a $\beta$-law:   
\begin{equation}
\vw (r') = v_{\rm inf} (1-R_0/r')^{\beta}, 
\label{eq:vw} 
\end{equation}
where $R_0$=$\rstar$$[1-(v_0/v_{\rm inf})^{1/\beta}]$, $\rstar$ is the radius of 
the supergiant star, $v_0/v_{\rm inf}$$\simeq$0.01, $v_{\rm inf}$ is the terminal wind velocity, and $r'$ is the radial distance from the supergiant star (such that $\vec{r'}$ is directed radially outward from the surface of the supergiant and $|\vec{r'}|$=$r$ at the NS location). The parameter $\beta$ describes the steepness of the wind velocity law. We assume for simplicity throughout our paper $\beta$=1, unless stated otherwise. The parameter $\beta$ is known to vary roughly between 0.8-2 depending on the specific type of supergiant star \citep[see, e.g.,][for a recent review]{nunez17} and, although it can affect significantly the profile of the stellar wind (and thus in turns the accretion process onto the NS), this effect is not qualitatively impacting the results presented in this paper. This is discussed more extensively in Sect.~\ref{sec:applications}. The terminal wind velocity is related to the escape velocity from the surface of the star 
\begin{equation}
v_{\rm esc} = \sqrt{\frac{2(1-\Gamma_{\rm e})G\mstar}{\rstar}}, 
\label{eq:vesc} 
\end{equation}
through the relation \citep{lamers95}:   
\begin{displaymath}
\frac{v_{\rm esc}}{v_{\rm inf}} \simeq \left\{ \begin{array}{ll}
2.6 & \textrm{if $T_{\rm eff}$ $>$21000~K}\\
1.3 & \textrm{if $T_{\rm eff}$ $\simeq$21000~K}\\
0.7 & \textrm{if $T_{\rm eff}$ $<$10000~K}. 
\end{array} \right.
\end{displaymath}
In Eq.~\ref{eq:vesc}, $\Gamma_{\rm e}$=$(\sigma_{\rm e}\lstar)$/$(4\pi c G\mstar)$, 
$\sigma_{\rm e}$$\simeq$0.30~cm$^2$/g, $\lstar$ is the bolometric luminosity of the supergiant companion, and $M_{\star}$ its mass loss rate. $\Gamma_{\rm e}$ is taking into account the correction from the radiation pressure due to electron scattering. The bolometric luminosity is also related to the mass loss rate of the star through the empirical relation: 
\begin{equation}
\log_{10}[\dotm]=-1.37+2.07\log_{10}[\lstar/(10^6\lsun)]-\log_{10}[v_{\rm inf}\rstar^{0.5}] 
\end{equation}
where $\lsun$, $\msun$, $\rsun$ are the luminosity, mass, and radius of the Sun, respectively; 
$\dotm$ is in units of $\msun$/yr, $v_{\rm inf}$ in km/s, $\rstar$ in $\rsun$, and $\dotm$ is the 
mass loss rate of the star. Note that formally these equations are valid only for O and B supergiant stars, and other prescriptions have been published in the literature \citep[see, e.g.,][and references therein]{vink2000,puls08}. Here, we followed a well know prescription for consistency within our calculations and we verified that slightly updated or different prescriptions would not significantly affect our results. For all objects of interest for this paper (see Sect.~\ref{sec:results}) the uncertainties on the measured OB supergiant parameters is far larger than any uncertainty introduced by the different prescriptions available to estimate the stellar mass loss rate.

\subsection{Simplified wind accretion scenario}
\label{sec:simpler}  

In this first paper, we limit our semi-analytical calculations to the case of the so-called Bondi-Hoyle spherical accretion scenario \citep{bondi52}. We follow, in particular, the treatment by \citet{davies79} and define first the accretion radius R$_{\rm a}$ as the distance from the NS at which the inflowing matter is gravitationally focused toward the compact object: \begin{equation}
R_{\rm a}=2GM_{\rm NS}/v_{\rm rel}^2.  
 \label{eq:ra}
\end{equation}
As anticipated in Sect.~\ref{sec:orbit}, we do not neglect\footnote{Note that in Eq.~\ref{eq:ra} we are still assuming that $v_{\rm rel}$$\gg$$c_{\rm s}$, where $c_{\rm s}$ is the local sound velocity. We verified that this condition is always satisfied in all cases treated in this paper.} the effect of the orbital velocity in the calculation of $v_{\rm rel}$. In the standard wind accretion theory it is assumed that all material from the companion star that enters into $R_{\rm a}$ can be accreted by the NS \citep[i.e., there is no consideration of the gating effects and/or the settling accretion process; see, e.g.,][]{illarionov75,bozzo08,shakura12}. If we assume that the supergiant wind is spherically symmetric, then the mass accretion rate within $R_{\rm a}$ is 
\begin{equation}
\dot{M}_{\rm acc} = \pi R_{\rm a}^2 \rho_{\rm w} v_{\rm rel} = \frac{1}{4} \left(\frac{R_{\rm a}}{r} \right)^2 \frac{v_{\rm rel}}{v_{\rm w}} \dot{M}_{\rm w} 
\label{eq:macc}
\end{equation}
where we used for the wind density $\rho_{\rm w}$ 
\begin{equation}
\rho_{\rm w} = \frac{\dot{M}_{\rm w}}{4\pi r^2 v_{\rm w}}. 
\end{equation}
The X-ray luminosity released by the wind accretion onto the compact object can be estimated from Eq.~\ref{eq:macc} by using the relation \citep[see, e.g.,][]{frank02}: 
\begin{equation}
L_{\rm acc} \approx \frac{G M_{\rm NS}}{R_{\rm NS}} \dot{M}_{\rm acc}. 
\label{eq:lacc} 
\end{equation}

\subsubsection{X-ray photoionization of the supergiant stellar wind}
\label{sec:ionization2} 

The X-rays emitted as a consequence of the accretion onto the NS in a wind-fed system can photoionize the stellar wind and cut-off the radiative force that is the main mechanism through which the wind is accelerated \citep{lamers99}. A detailed and analytical study of this effect was presented originally by \citet{ho87}.  We follow here their treatment, but improve it by removing the approximation $v_{\rm rel}$$\simeq$$v_{\rm w}$ and including the effect of eccentric orbits. 

It is assumed that the wind of the supergiant star is accelerated according to Eq.~\ref{eq:vw} up to a distance $r_{\xi}$ from the star (the so-called Str$\ddot{\rm o}$mgren sphere) at which the X-rays coming from the NS photoionizes the wind material and halt the 
acceleration \citep[see, e.g.,][and references therein]{ho87,manousakis12}. From this distance onward (in the direction of the NS), the wind material proceeds with a constant velocity according to the equation: 
\begin{displaymath}
v_{\rm w}= \left\{ \begin{array}{ll}
v_{\rm inf} (1-R_0/r)^{\beta} & \textrm{if $r$$=$$r_{\xi}$}\\
v_{\rm inf} (1-R_0/r_{\xi})^{\beta} & \textrm{if $r$$>$$r_{\xi}$}. 
\end{array} \right.
\end{displaymath}
The distance $r_{\xi}$ is defined by the equation
\begin{equation}
\frac{L_{\rm acc}}{n(r_{\xi})(r-r_{\xi})^2}=\xi_{\rm cr},
\label{eq:xicr} 
\end{equation}
where $\xi_{\rm cr}$ is the value of the ionization parameter $\xi$ at which the acceleration force is cut-off, $n(r)$$\simeq$$\rho(r)$/$m_{\rm H}$, and $m_{\rm p}$ is the proton mass \citep[see][]{tarter69}. 
The value of $\xi_{\rm cr}$ is not well known and different authors have assumed largely different values in the range 1-10$^{4}$~erg~cm~s$^{-1}$ \citep[see, e.g.,][and references therein]{ducci10,krtika15,mellah18}. Using together Eq.~\ref{eq:macc}, \ref{eq:lacc}, and \ref{eq:xicr}, we obtain: 
\begin{equation}
\xi_{cr}=4\pi m_{\rm H} \frac{G^3 M_{\rm NS}^3}{R_{\rm NS}} \frac{1}{r^2\left(\frac{r}{r_{\xi}}-1\right)^2}\frac{1}{v_{\rm rel}^3(r_{\xi})}.  
\label{eq:feedback} 
\end{equation}
This is the generalized form of the steady-state equation presented by \citet[][see their equation 6]{ho87}. It can be solved to derive the distance $r_{\xi}$ from the supergiant at which the X-ray radiation from the compact object is able to halt the acceleration of the companion star wind, in turns regulating the mass accretion rate and the production of the X-ray emission itself. 
In Sect.~\ref{appendix:ho87} we show that this form of the feedback equation might have up to three solutions for $r_{\xi}$, given a fixed set of the free parameters. However, only one at a time fulfills all physical boundary conditions. Given the present uncertainties on the ionization parameter and the large range of values assumed by different authors in the literature, we investigate in the following sections the effect of different assumptions on $\xi_{\rm cri}$ on the presented results. 

Depending mainly on the mass loss rate from the supergiant star and the orbital separation, it is possible to have cases where Eq.~\ref{eq:feedback} predicts a Str$\ddot{\rm o}$mgren sphere extending up to the surface of the supergiant (see details in appendix~\ref{appendix:ho87}). This aspect is discussed more extensively in Sect.~\ref{sec:off}. We note that \citet{blondin94} introduced the concept of a thermally driven wind in SgXBs, i.e. a stellar wind which emission originates by the energy deposition of the X-rays from the NS onto the supergiant atmosphere. According to this paper, the thermally excited wind would take the place of the radiatively accelerated wind in systems where the Str$\ddot{\rm o}$mgren sphere extends to the surface of the supergiant. However, at the best of our knowledge, there have been no observational evidences of thermally excited winds in SgXBs and thus do not discuss further this possibility.

\subsubsection{Avoiding the formation of an accretion disk around the NS}
\label{sec:disk2} 

An accretion disk around the NS in a binary system with a massive companion can be formed if: (i) the distance between the compact object and the massive companion is sufficiently small for the latter to fill its Roche Lobe, or (ii) the angular momentum of the stellar wind material flowing across the NS accretion radius is sufficiently large \citep[see, e.g.,][]{frank02}.  

Concerning (i), we assume for all the cases of interest here that the ratio between the mass of the NS and the supergiant ($q$=$M_{\rm NS}$/$M_{*}$) is 0.03$<$$q$$<$1 and thus the Roche Lobe of the massive star can be estimated as \citep{frank02}:  
\begin{equation}
R_{\rm L*}\simeq\left(\frac{M_*}{M_{\rm NS}}\right)^{0.45} R_{\rm LNS}
\end{equation}
where $R_{\rm LNS}$=a-$R_{\rm L*}$ is the NS Roche Lobe. As the on-set of the Roche-Lobe overflow would cause the system to switch from being wind-fed to disk-fed, thus drastically changing the physics of the accretion process and invalidating our treatment in Sect.~\ref{sec:simpler}. Here, we simply verify that in all cases of interests for this paper the condition $R_{*}>R_{\rm L*}$ never occurs (see Sect.~\ref{sec:results}).  

Equivalently, we have to ensure that also the condition in (ii) never occurs for all systems of interest discussed in Sect.~\ref{sec:results}. To evaluate the conditions necessary for the formation of an accretion disk in the case (ii), we follow the argument presented by \citet{wang81} and also adopted by \citet{ducci10}. These authors have showed that the wind matter entering the NS accretion radius from opposite sides can be endowed with a residual angular momentum if azimuthal perturbations in the wind density and velocity are induced by the presence of the NS. Assuming reasonable amplitudes for these perturbations, it has been shown that the amount of residual angular momentum that would allow the wind material to settle into a keplerian orbit within the accretion radius con only be achieved for very slow winds. In particular, it turns out that the wind velocity at the NS location has to be of the same order of the NS orbital velocity (such that they nearly cancel out). Although the analytical treatment discussed by these authors is endowed with large uncertainties, it is interesting to note that the same result is achieved by a more recent publication, in which the formation of an accretion disk within the accretion radius of a NS in a HMXB is investigated by means of 3D hydrodynamic simulations \citep{mellah19}. Comparable results were suggested by \citet{karino19} \citep[see also][]{xu19}. For the goals of this paper, we thus consider that a SgXB can switch from wind to disk accretion when $v_{\rm rel}\simeq$0, and thus that this is also the limit of validity of our calculations.

\section{Results}
\label{sec:results}

\subsection{Solutions to the generalized equation of the X-ray feedback on the stellar wind}
\label{appendix:ho87}  

In this section we analyze the solutions to the generalized equation of the stellar wind photoionization given in Sect.~\ref{sec:ionization2} and compare the new solutions with the previous ones published in the literature. 

In general terms, the solutions to Eq.~\ref{eq:feedback} provide an estimate of the distance between the NS and the massive companion at which the X-ray irradiation 
emitted by the compact object is able to photoionize the wind material and turn off the main acceleration mechanism of the stellar wind. As in the present case we generalized the equation to the case of eccentric orbits, the solutions to Eq.~\ref{eq:feedback} are functions of the orbital phase. In order to investigate how the solutions change compared to those previously published in the literature, let's first rewrite Eq.~\ref{eq:feedback} as
\begin{equation}
f(x)=0,
\label{eq:fx} 
\end{equation} 
where
\begin{equation}
f(x)= k \frac{1}{x_0^2\left(\frac{x_0}{x}-1\right)^2}\frac{1}{v_{\rm rel}^3(x)} -1,    
\end{equation}
$x_0$=$r$/$R_0$, $x$=$r_{\xi}$/$R_0$, and k contains all numerical constants:
\begin{equation}
k = 4\pi m_{\rm H} \frac{G^3 M_{\rm NS}^3}{R_{\rm NS}} \frac{1}{\xi_{cr} R_0^2}.     
\end{equation}

\begin{figure*}
\centering
\includegraphics[scale=0.35]{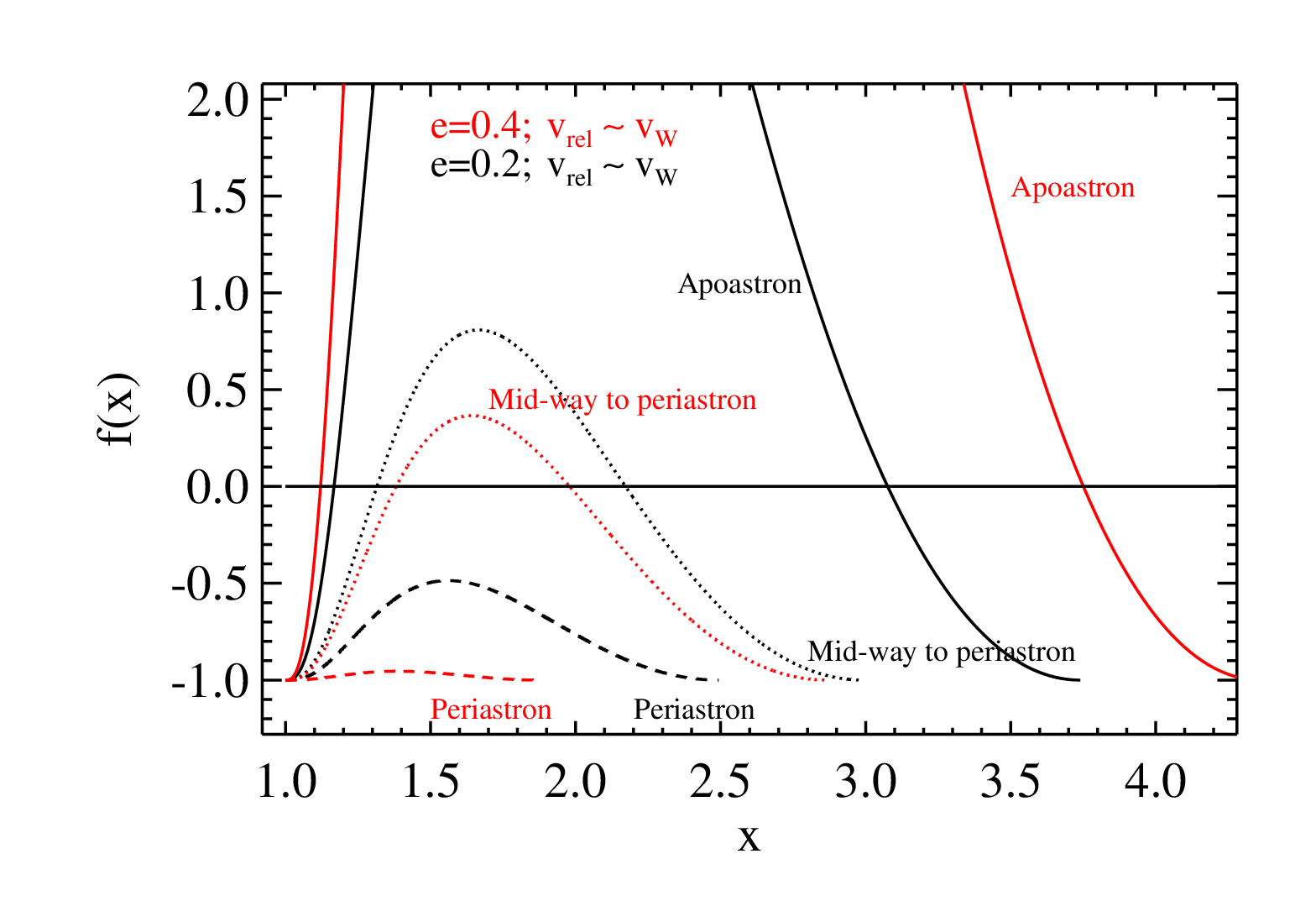}
\includegraphics[scale=0.35]{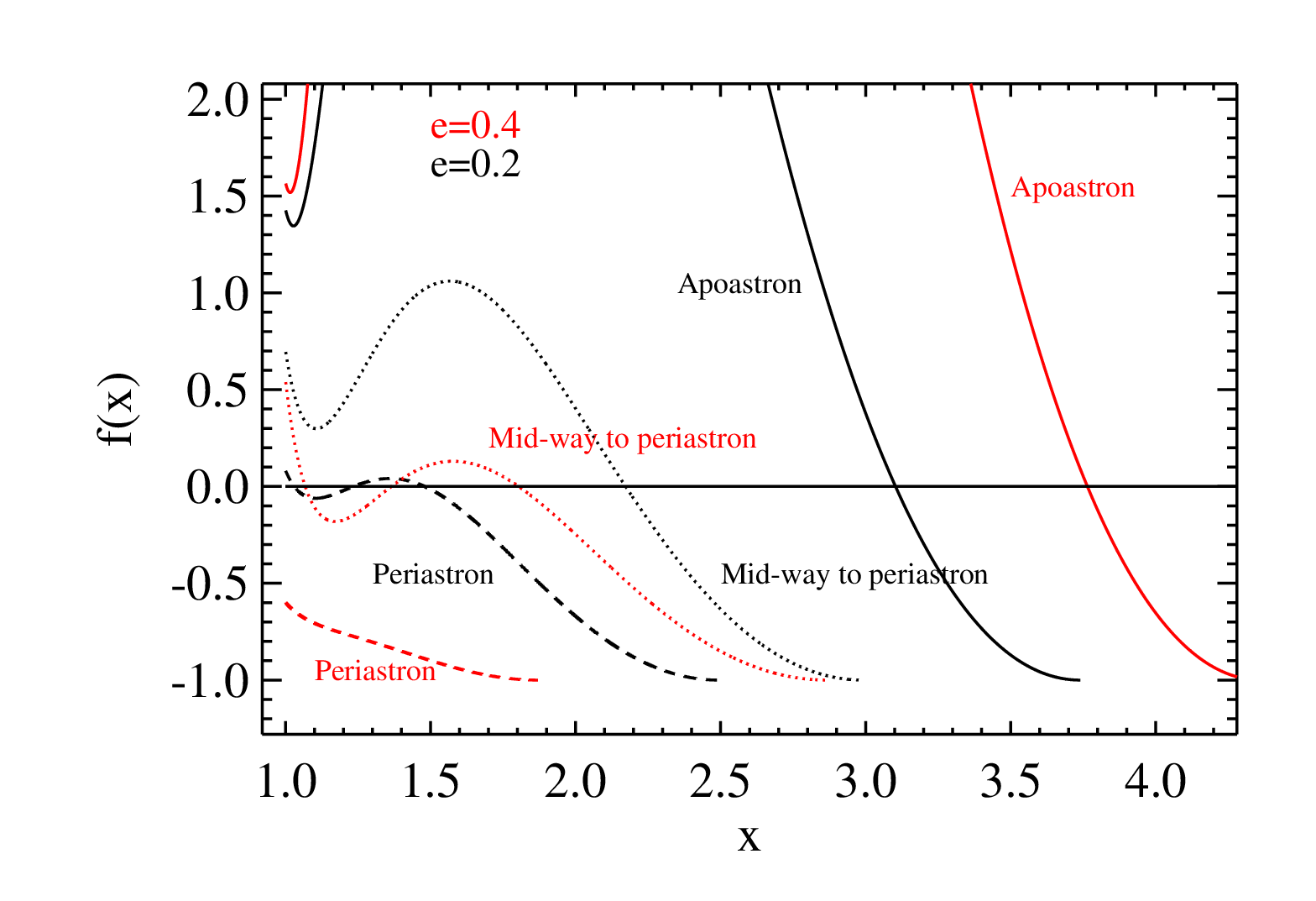}
\includegraphics[scale=0.35]{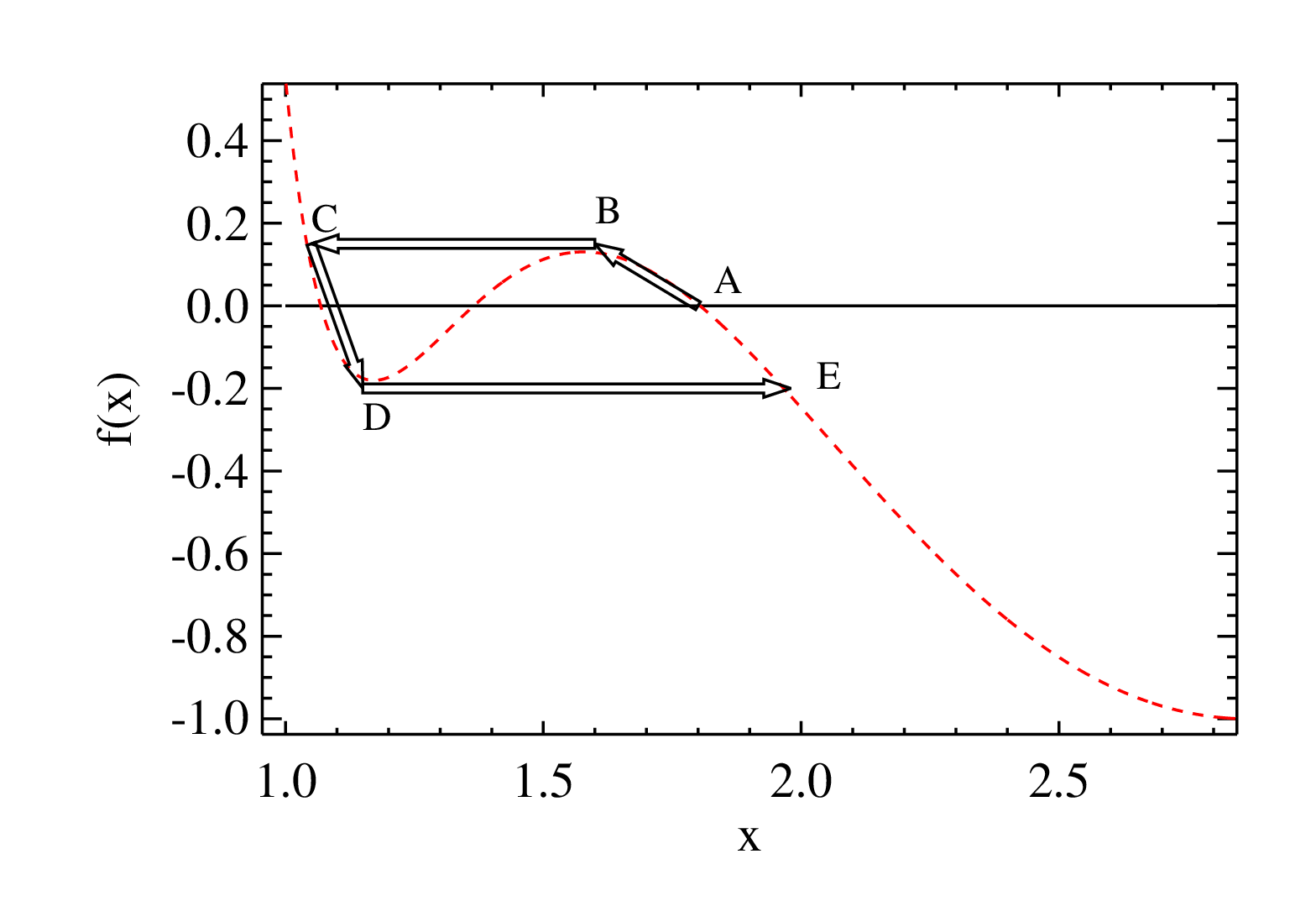}
\includegraphics[scale=0.35]{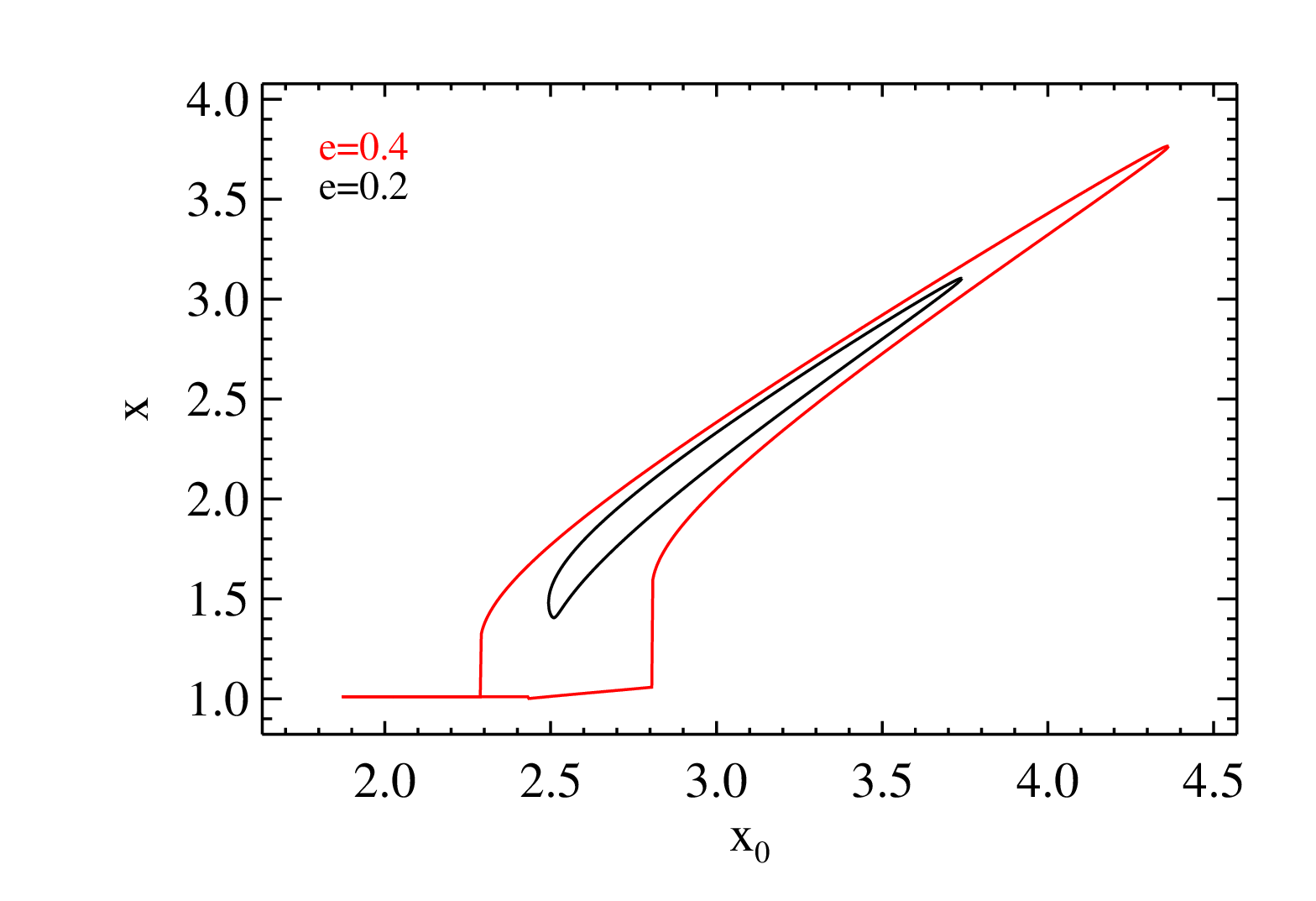}
\caption{{\it Top left}: Plot of $f(x)$ for different values of $x_0$ when the approximation $v_{\rm rel}$$\simeq$$v_{\rm w}$ is used. The system parameters assumed here are the same as in Fig.~\ref{fig:coord}. The black lines correspond to an eccentricity for the NS orbit of $e$=0.2, while the red lines are for $e$=0.4. In both cases the continuum line corresponds to the $f(x)$ calculated for the NS at the apoastron, the dotted line is mid-way toward the periastron, and the dashed line is for the NS at periastron. {\it Top right}: Same as before but without making use of the approximation $v_{\rm rel}$$\simeq$$v_{\rm w}$. {\it Bottom left}: Zoom of the red dashed line in the top right panel. We show the instability-like solutions to the feedback equation (see text for details). {\it Bottom right}: Ionization radius $x$=$r_{\xi}$/$R_0$ calculated for the same system parameters of the other 
figures. The black line corresponds to the case $e$=0.2, while the red line is for $e$=0.4.}  
\label{fig:solioniz} 
\end{figure*}

We consider first the approximated solutions with $v_{\rm rel}$$\simeq$$v_{\rm w}$ (i.e. neglecting the NS orbital velocity) discussed by both \citet{ho87} and \citet{karino14}. In the left panel of Fig.~\ref{fig:solioniz}, we show how the function $f(x)$ with the simplification above is changing while the NS moves from $x_0$=0 (apoastron) to $x_0$=$a$/$R_0$ (periastron) in relatively high eccentric orbits (we show two eccentricity cases, $e$=0.2 and $e$=0.4). When the NS is far away from the companion, two solutions can be found to the Eq.~\ref{eq:fx}. The first solution gives a radius of the Str\"omgren sphere ($r_{\rm \xi}$) very close to the surface of the supergiant companion ($x$$\simeq$1). This is the ``high luminosity'' solution identified by both \citet{ho87} and \citet{karino14}, and corresponds, according to their interpretation, to a situation in which the NS is able to accrete a conspicuous amount of material from the supergiant. The second solution (``low luminosity solution'') would give instead  $x$$\gg$1 and corresponds to a situation 
in which the luminosity released from the accretion is significantly lower, and only a small fraction of the wind material can be ionized close to the source of the X-ray radiation (solutions are searched in all cases only up to $x$$\lesssim$$x_0$; the value of $x_0$ decreases when the NS moves from apoastron to periastron, explaining 
the different extensions of the curves along the x-axis). 
When the NS approaches the supergiant companion and moves close to the apoastron, $f(x)$ assumes only negative values and Eq.~\ref{eq:fx} has no solution. The reason for this behavior is that, when the NS gets too close to the companion, the X-ray luminosity produced by the enhanced accretion is very high and the wind material is easily ionized down to the surface of the supergiant star (for the set of parameters used here). The top left panel of Fig.~\ref{fig:solioniz} also shows that this behavior is more evident at higher 
eccentricities, when the distance between the NS and the companion at periastron is smaller. \citet{ho87} commented that both the low and high luminosity solutions had to be consider to reconcile the theoretically estimated X-ray luminosity of several wind accreting systems with the value inferred from observations. The existence of systems in one or the other regime could be due, according to their interpretation, to evolutionary effects. \citet{karino14} suggested that the high luminosity solution could explain the classical SgXBs, while the low luminosity solution is useful to explain the SFXT sources. However, there is not a clear argumentation of what should make one or the other branch of solution applicable to different classes of SgXBs and thus this conclusion remained so far speculative. 
   
By removing the approximation $v_{\rm rel}$$\simeq$$v_{\rm w}$, we argue that a different scenario and interpretation can be proposed for the solutions to the Eq.~\ref{eq:fx}. This is shown in the right panel of Fig.~\ref{fig:solioniz}. The curves represented here are the same as in the previous case, but we included the orbital velocity in the estimate of $v_{\rm rel}$ using the treatment in Sect.~\ref{sec:orbit}. When the NS is far away from the companion, the X-ray emission marginally affects the wind of the massive star and a first solution to the equation Eq.~\ref{eq:fx} appear at $x$$<$$x_0$. Closer to the supergiant, the solution to $f(x)=0$ reaches one of the minima of $f(x)$ and, apparently, up to three solutions are possible for each value of $x$. These multiple solutions are found until a critical distance of the NS from the companion is reached at which the solution to Eq.~\ref{eq:fx} corresponds to the maximum of $f(x)$. For separation distances smaller than this critical value only a single solution 
to Eq.~\ref{eq:fx} is found for $x$$\simeq$$x_0$. These solutions correspond to the case in which the accretion luminosity is so high that most of the wind material is ionized and the  Str\"omgren sphere extends to the surface of the supergiant star. 

At odds with the work by \citet{ho87}, the calculations developed in the present work permit us to follow the evolution of the different solutions of the Eq.~\ref{eq:feedback} along the NS orbit, i.e. as a function of the separation between the compact object and the supergiant star. As the apparent multiple solutions can be found only when the NS gets closer and closer to the companion, continuity arguments suggest that these different solutions can not co-exist at the same time but rather than we are facing an instability-like problem. This is explained graphically in the bottom left panel of Fig.~\ref{fig:solioniz}. We argue that, while the NS is moving from apoastron to periastron, stable solutions for the ionization problem moves from point ``A'' to point ``B''. At this stage, when the NS arrives at a critical distance from the companion, no stable solutions can be found: the X-ray irradiation is far too high and all the wind material is completely photoionized up to the supergiant surface. As this ionization cut-off the main acceleration mechanism of the stellar wind (see Sect.~\ref{sec:intro}), the mass loss rate from the star is strongly suppressed. The solution thus jumps from point ``B'' to point ``C''. As in this situation there is no wind that can produce accretion (and thus further X-rays), lacking any other viable accretion mechanism, the source of the X-ray emission is largely reduced and the radius of the Str\"omgren sphere rapidly decreases 
and moves on local dynamical times scale from ``C'' to ``D''. Once the NS and the companion star reached a sufficiently large orbital separation, the accretion luminosity is low enough to get a new stable solution. This marks the passage from ``D'' to ``E''. The resulting value of the Str\"omgren sphere is plot in the bottom right panel of Fig.~\ref{fig:solioniz}. Here it can be noticed that only for the higher value of the eccentricity ($e$=0.4, given all other parameters fixed for these calculations) the NS gets close enough to the supergiant to ionize completely all wind material down to its surface. The nearly vertical drops in the red line show the points along the NS orbit where the separation between the two stars become sufficiently close for this high ionization to take place. Close to these positions, the value of the ionization radius rapidly approaches the supergiant surface ($x$=$r_{\xi}$/$R_0$$\simeq$1). Note that, being the relative velocity between the NS and the companion asymmetric during the passage from apoastron to periastron and vice-versa (see Fig.~\ref{fig:coord}), the values of $x$ in the bottom right panel of Fig.~\ref{fig:solioniz} draw different elliptical shapes for each value of the eccentricity. 

Our conclusions thus show that there is not a bi-modality in the accretion modes of SgXBs (and other wind-fed HMXBs) as assumed by \citet{ho87}, \citet{krtika12}, and \citet{karino14}. It is rather that a wind-fed HMXB can experience both kinds of low and high luminosity states at different times along one orbital revolution. The switch between these states is caused by the lack of stable solutions to the equation of the feedback (Eq.~\ref{eq:feedback}) when the photoionization gets high enough for the Str\"omgren sphere to approach the surface of the massive companion (the branch of solutions between points C and D in the bottom left panel of Fig.~\ref{fig:solioniz}). This thus provides the missing physical explanation and a context for the applicability of the low and high luminosity solutions that were already identified decades ago for SgXBs.

\subsubsection{Off-states in SgXBs}
\label{sec:off}

\begin{figure*}
\centering
\includegraphics[scale=0.35]{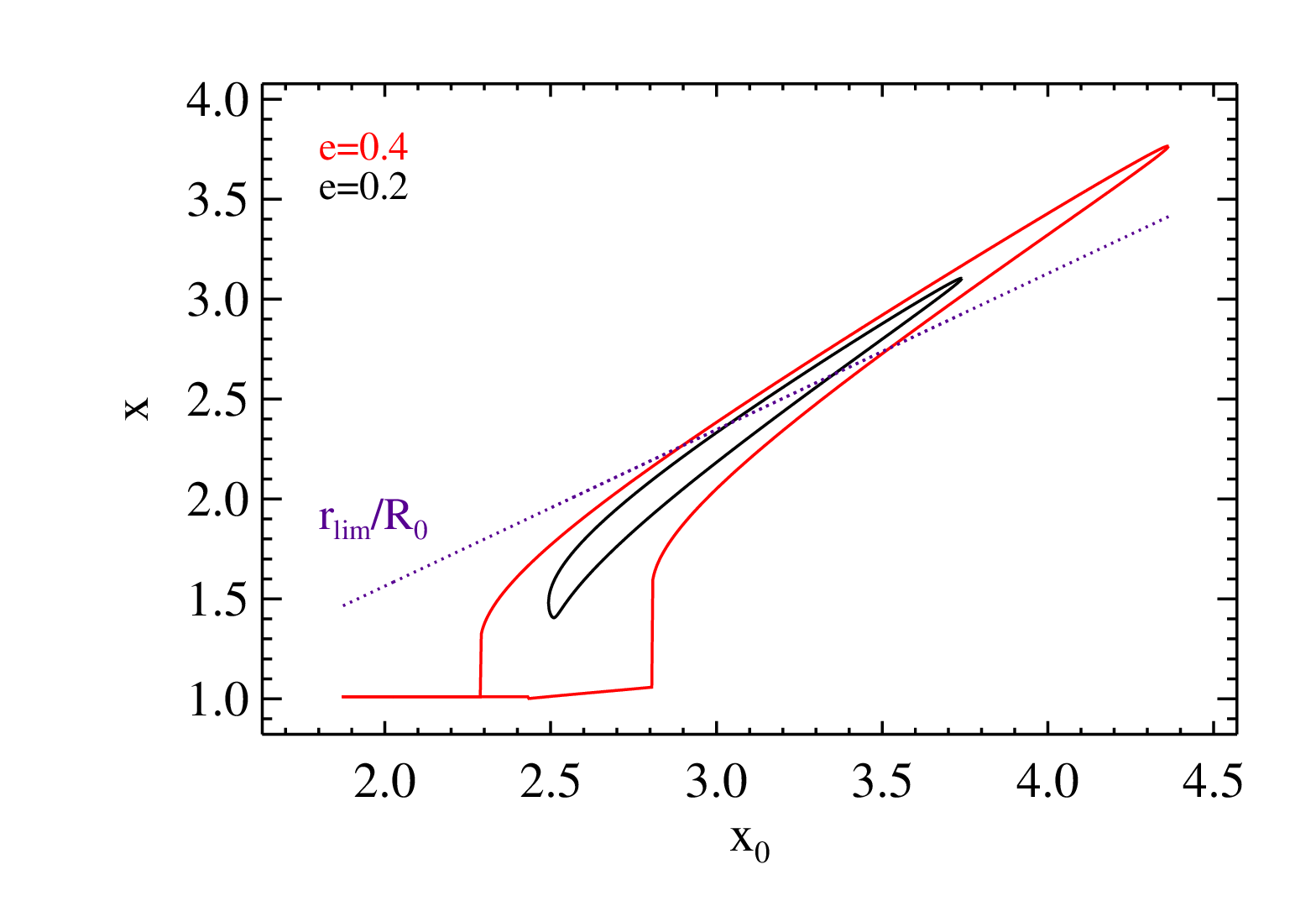}
\includegraphics[scale=0.35]{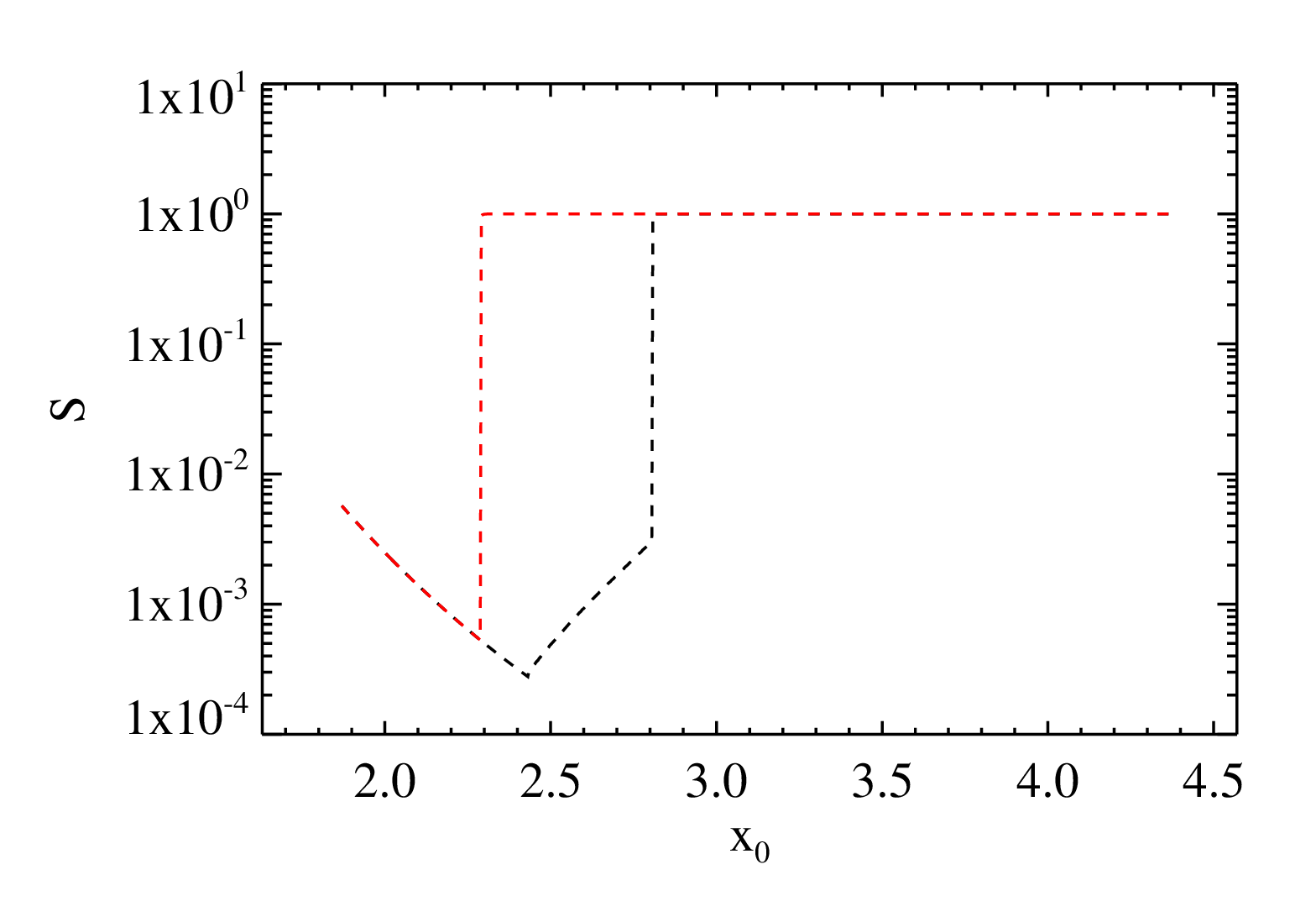}
\caption{{\it Left}: Same as lower right panel in Fig.~\ref{fig:solioniz}, but we indicated here also the value of $r_{\rm lim}$ (dotted violet lines) in units of the stellar radius. {\it Right}: Corresponding values of the density probability $S$. For the eccentricity $e$=0.2 
$S$$\simeq$1 for all values of $x_0$ and thus this case is not shown in the figure. For the eccentricity $e$=0.4, $S$ significantly differs from unity when $r_{\xi}$$\simeq$$R_0$, i.e. $x$$\simeq$1. In this figure, 
the red dashed line corresponds to the left upper branch of values of $x$ in the left panel (upper left half of the deformed ellipse), while the black dashed line is for the lower right branch (lower right half of the deformed ellipse).}  
\label{fig:solioniz2} 
\end{figure*}

We also argue that in those cases where the system X-ray luminosity is high and the photoionization proceeds closer and closer to the supergiant star, accretion is partly inhibited due to the limited acceleration of the wind. This can be proved by using a modified version of the argument presented by \cite{ducci10}. Once the radiation force has been cut-off by the ionization effect of the X-rays (at a distance from the primary star $r$$\geq$$r_{\xi}$), the wind material is no longer accelerated. This material is subjected only to the gravitational forces imposed by the compact object and the supergiant star. The latter should also be diminished by the action of the electron scattering. There is thus a value of the distance from  the supergiant star, $r_{\lim}$$\geq$$r_{\xi}$, at which these forces balance each other, i.e. 
\begin{equation}
\frac{G M_{*}}{r_{\rm lim}^2}(1-\Gamma_{\rm e})=\frac{G M_{\rm NS}}{(r-r_{\rm lim})^2}, 
\end{equation}  
where $\Gamma_{\rm e}$=$\sigma_{\rm e}$$L_{*}$/(4$\pi$$c$$G$$M_{*}$), and the opacity for electron scattering is $\sigma_{\rm e}$$\simeq$0.3\,cm$^2$\,g$^{-1}$. 
As we are interested only in those solutions for which $r_{\rm lim}$$\leq$r, we obtain: 
\begin{equation}
r_{\lim}=\frac{J}{J-1} r \left(-1+\frac{1}{\sqrt{J}}\right), 
\end{equation}
where $J$=$M_{*}$/$M_{NS}$(1-$\Gamma_{\rm e}$) and $J$$\simeq$10 for typical values used in this paper. In all those cases in which $r_{\rm lim}$$>$$r_{\xi}$, we need to check if particles from the wind accelerated up to $r_{\xi}$ are endowed with a sufficiently high velocity to reach the NS. In practice, accretion in these conditions is possible only if $v_{\rm w}(r_{\rm lim})$$>$0. Assuming that a particle arrives at $r_{\rm lim}$ 
with null velocity, we can integrate backward the momentum equation of a particle to find the minimum velocity $v_{\rm w}(r_{\xi}$) below which ionization 
of the wind material not only cut-off the acceleration mechanism of the wind but also inhibits  accretion. We indicate with $R$ the separation between the NS and the companion, and write the momentum equation of the wind particle located between $r_{\xi}$ and $r_{\rm lim}$ as:  
\begin{equation}
v\frac{dv}{dR} = - \frac{GM_{*}}{R^2}(1-\Gamma_{\rm e})+\frac{GM_{NS}}{(r-R)^2}.  
\end{equation}
The integration mentioned above gives: 
\begin{equation}
\int_{v_{\rm w}(r_{\xi})}^{0} v dv = - \int_{r_{\xi}}^{r_{\rm lim}} \left[\frac{GM_{*}}{R^2}(1-\Gamma_{\rm e})-\frac{GM_{\rm NS}}{(r-R)^2}\right] dR,
\end{equation} 
and thus the lowest possible value of the wind velocity at $r_{\xi}$ that would still allow significant accretion is 
\begin{eqnarray}
& v_{min}^2(r_{\xi}) = -2GM_{\rm NS} \left[ \frac{1}{r-r_{\rm lim}} - \frac{1}{r-r_{\xi}} \right] \nonumber \\
& -2GM_{*}(1-\Gamma_{\rm e})\left[\frac{1}{r_{\rm lim}}-\frac{1}{r_{\xi}}\right]. & 
\end{eqnarray}
Assuming that the wind particles have a Gaussian distribution of velocity centered on $v_{\rm w}$ (see Eq.~\ref{eq:vw}) with 
$\sigma_{\rm v}$$\simeq$10$^{7}$~cm~s$^{-1}$, the density probability of having particles at $r_{\xi}$ with a velocity 
larger than $v_{min}(r_{\xi})$ is 
\begin{equation}
S = \frac{1}{\sqrt{2\pi}\sigma_{\rm v}}\int_{v_{min}(r_{\xi})}^{\infty} \exp\left[-\frac{(v-v_{\rm w}(r_{\xi}))^2}{2\sigma_{\rm v}^2}\right] {\rm dv}. 
\end{equation}
This gives: 
\begin{equation}
S = \frac{1}{2} Erfc\left[\frac{v_{min}(r_{\xi})-v_{\rm w}(r_{\xi})}{\sqrt{2}\sigma_{\rm v}}\right], 
\label{eq:s}
\end{equation}
where $Erfc$=1-$Erf$ is the complementary of the error function. The function 1/2$Erfc(z)$ goes to unity very rapidly when $z$$\lesssim$2 and goes quickly to zero 
when $z$$\gtrsim$2. We verified that generally $v_{min}(r_{\xi})$$\ll$$v_{\rm w}(r_{\xi})$, and thus S$\simeq$1. However, this does not apply to those cases in which 
$r_{\xi}$$\simeq$$R_0$. In these cases S$\ll$1, thus indicating that in the high luminosity solutions the effective accretion can be much less ($\sim$10$^{-2}$-10$^{-4}$) 
than expected. This is illustrated in Fig.~\ref{fig:solioniz2}. 

We conclude from this analysis that the so-called high luminosity solutions correspond to short-lived states in which the temporarily high luminosity switches off the wind acceleration and the accretion with it. We should expect that, when a NS approaches its massive companion and the X-ray luminosity raises above the critical level to photoionize completely the wind up to the surface of the supergiant, the accretion is also inhibited and the X-ray luminosity is consequently also turned down. The source enters an ``off-state''. On local dynamical timescales, the disappearance of the ionizing X-ray flux causes immediately after the wind to be launched again, in turns restoring accretion while the NS moves away from the companion. We thus expect that a rough estimate of the duration of an off-state is: 
\begin{equation}
\tau_{\rm dyn}\simeq r/v_{\rm w} \sim R_0/v_{\rm w}. 
\end{equation}
For the typical parameters considered in this paper $\tau_{\rm dyn} \sim 2\times 10^3\,{\rm s}$. 
The presence of off-states has been observed in several SgXBs. We discuss more this point with an application to a few specific sources in Sect.~\ref{sec:applications}. 
\begin{figure*}
\centering
\includegraphics[scale=0.35]{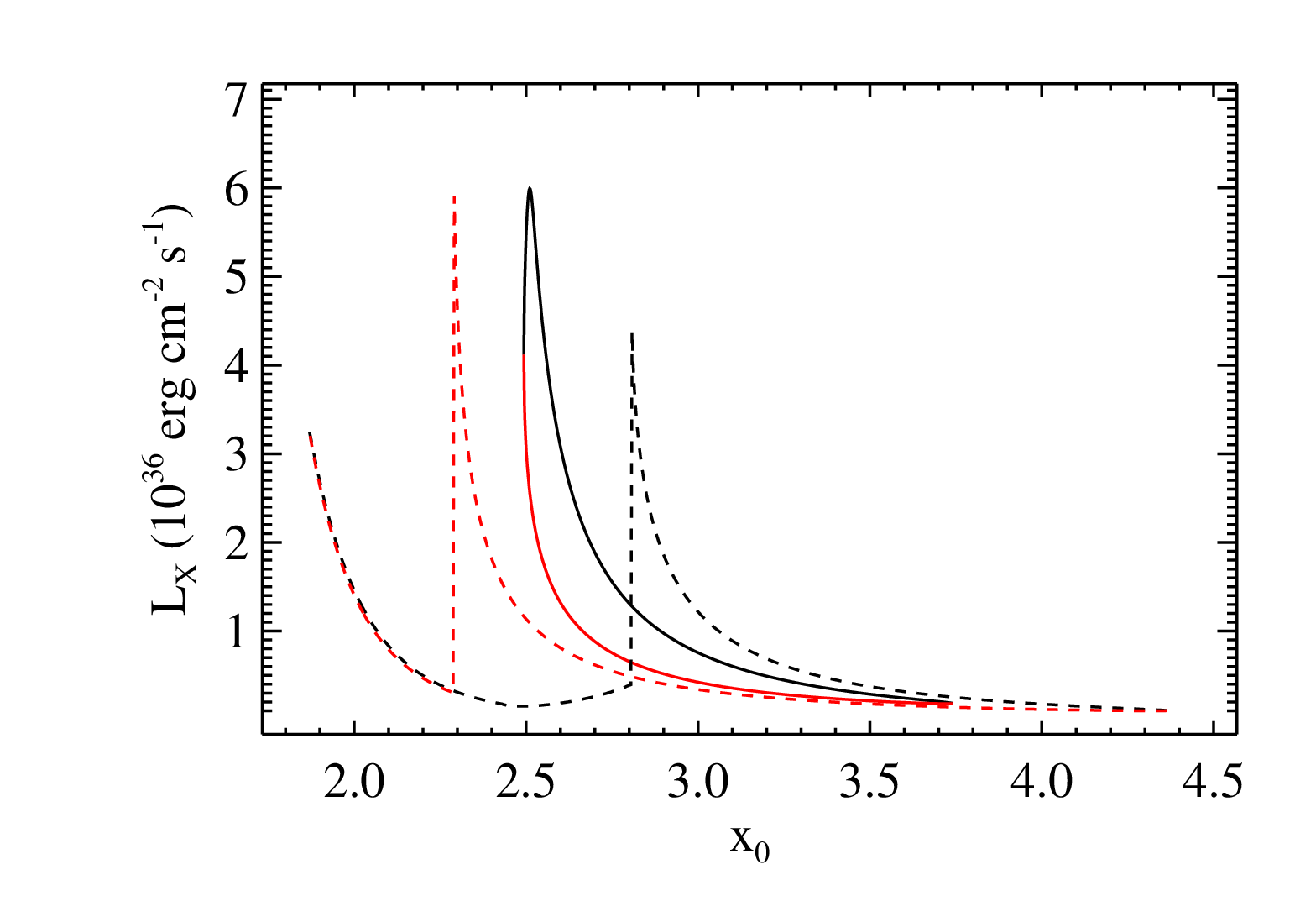}
\includegraphics[scale=0.35]{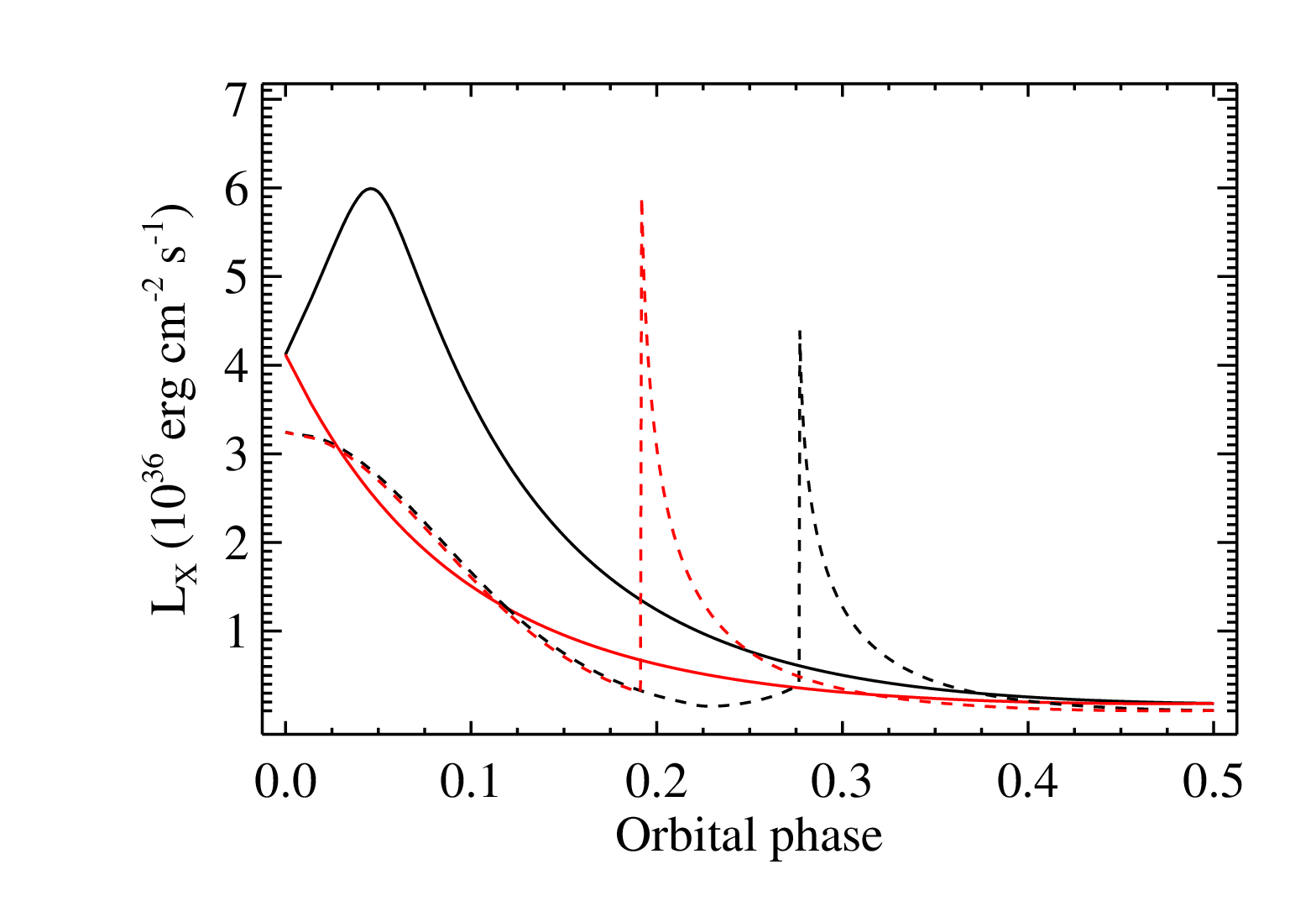}
\includegraphics[scale=0.35]{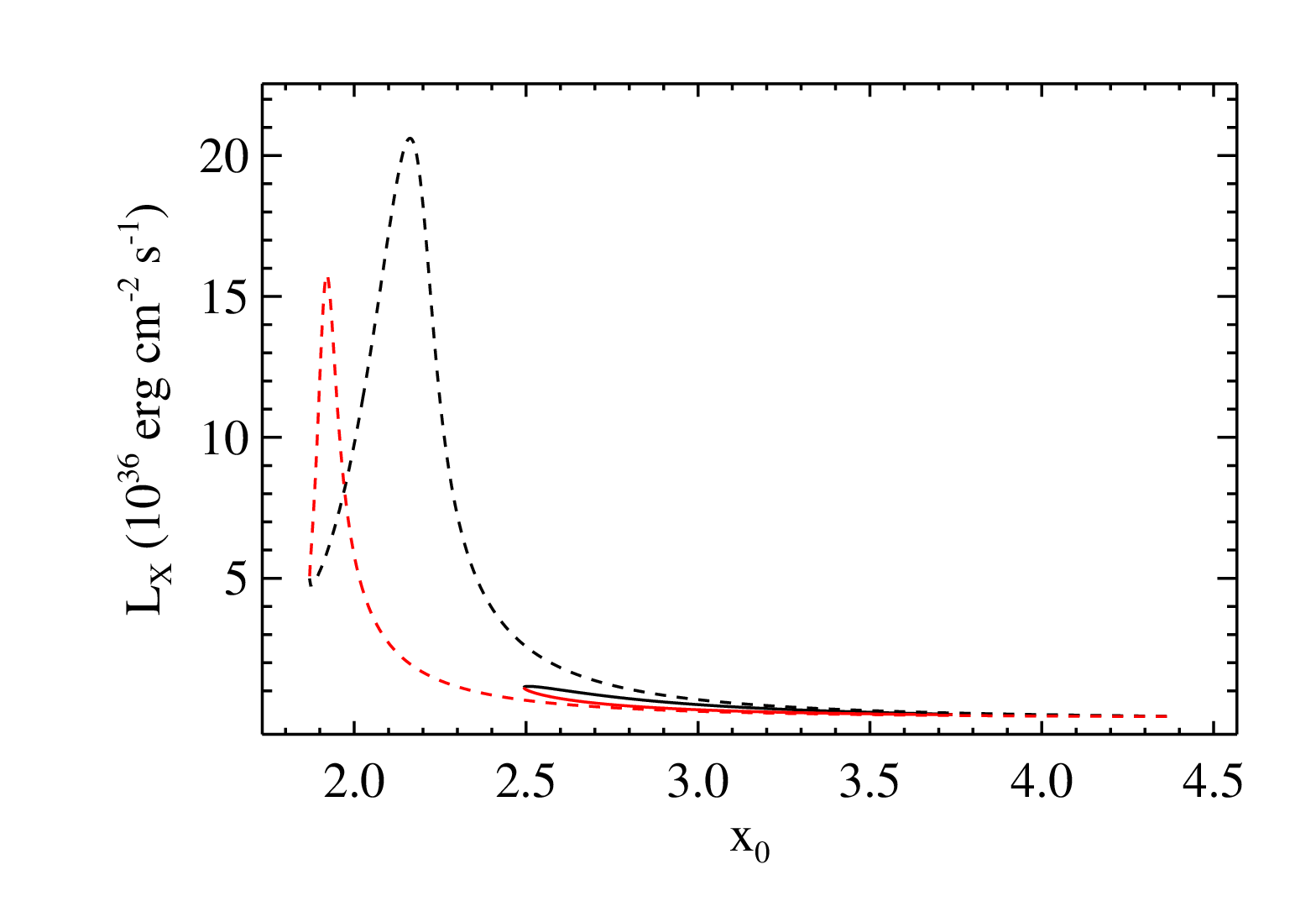}
\includegraphics[scale=0.35]{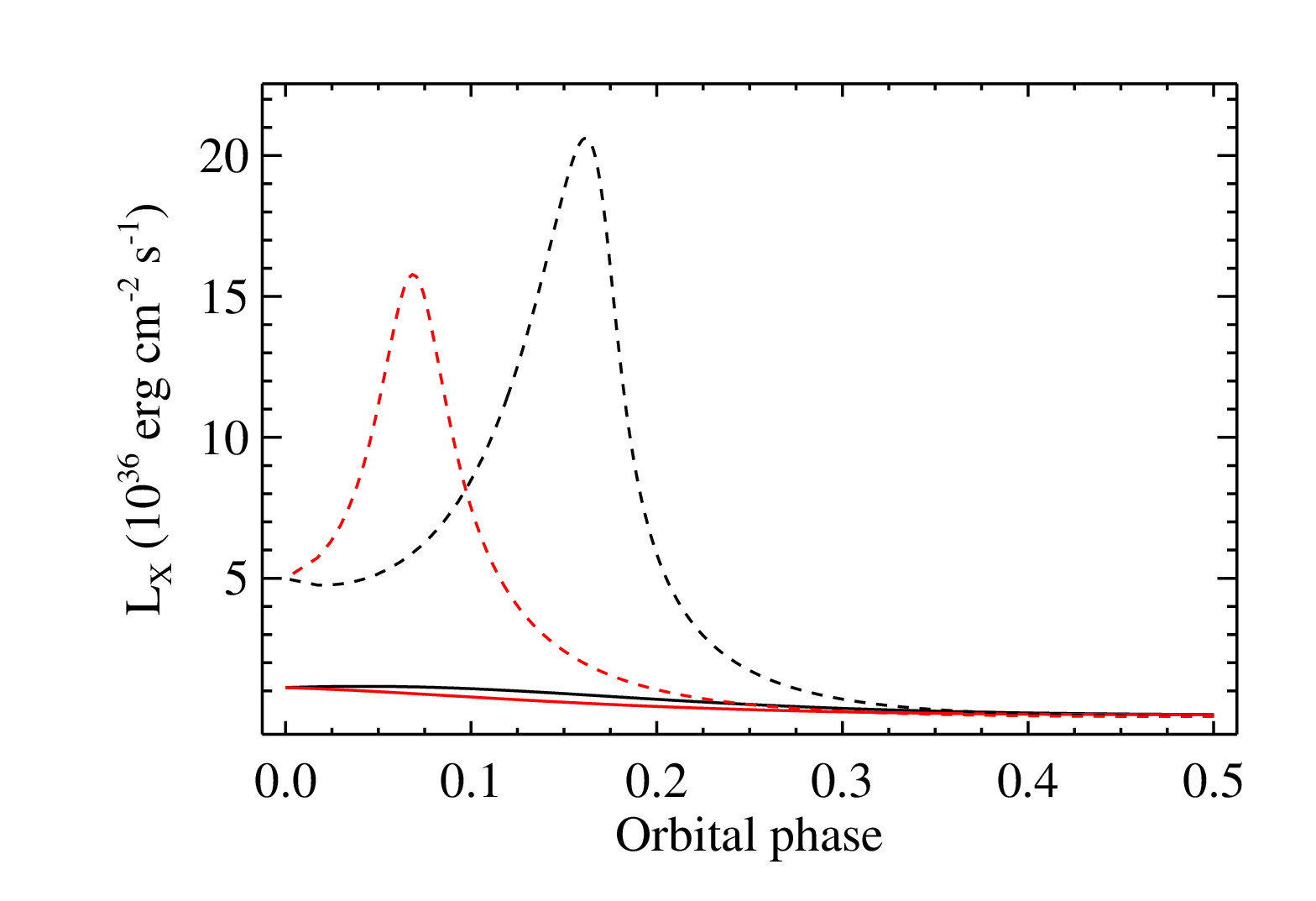}
\includegraphics[scale=0.35]{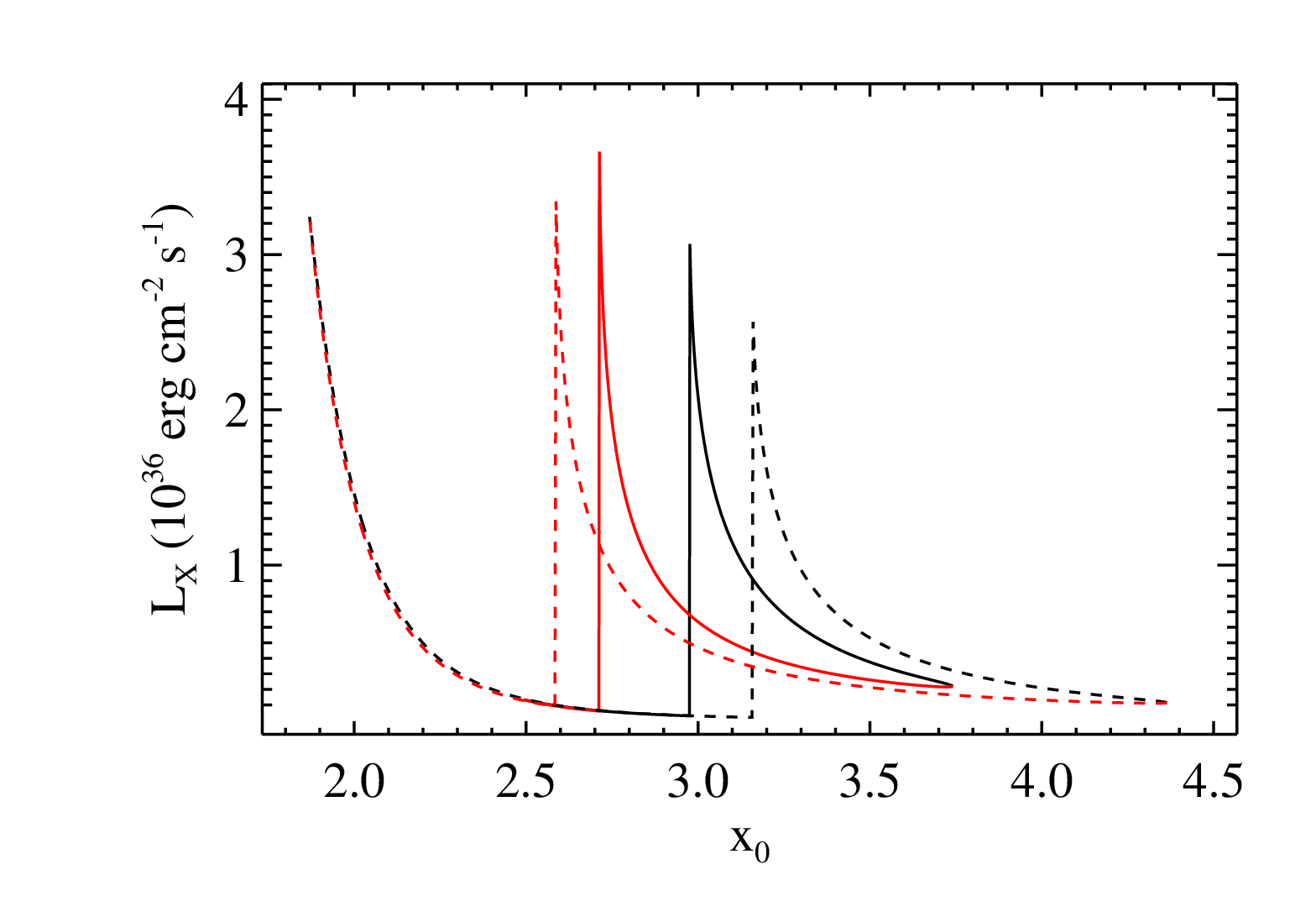}
\includegraphics[scale=0.35]{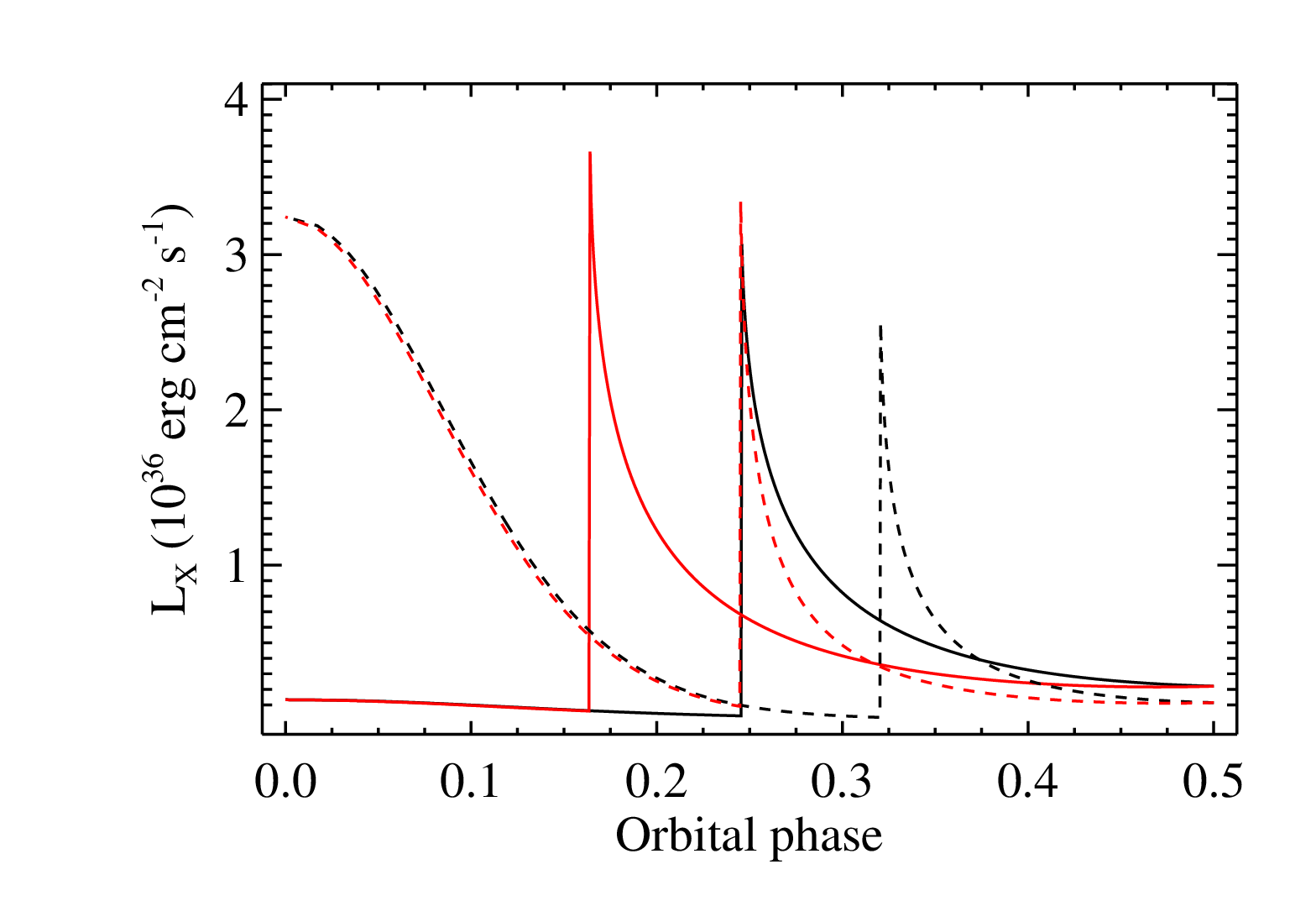}
\caption{{\it Top figures}: calculated variation of the X-ray luminosity along the orbit of a SgXB with the same parameters as those listed in the caption of Fig.~\ref{fig:coord}. The figure on the left shows the luminosity as a function of the orbital separation $x_0$ (in units of the stellar radii), while the figure on the right shows the luminosity as a function of the orbital phase as obtained from Eq.~\ref{eq:erre} (phase 0 is the periastron passage). We show in both figures the case of $e=0.2$ with solid lines and the case $e=0.4$ with dashed lines. The black lines correspond to the part of the orbit where the NS is moving away from the companion going from periastron toward apoastron. The red lines correspond to the phases of the orbit where the NS moves from the apoastron toward the supergiant companion. {\it Middle figures}: same as top figures but assuming $\xi_{\rm cri}$=500. {\it Bottom figures}: same as top figures but assuming $\xi_{\rm cri}$=50.}  
\label{fig:lxacc} 
\end{figure*}

\subsubsection{The accretion luminosity along the orbit}
\label{sec:lxacc}

Under the assumptions described in Sect.~\ref{sec:simpler}, we compute here as an example  the orbital lightcurve of a generic SgXB whose system parameters are identical to those listed in the caption of Fig.~\ref{fig:coord}. This example allows us to illustrate the effect of eccentric orbits convolved with the photoionization of the stellar wind. 

We show the outcomes of the calculations in Fig.~\ref{fig:lxacc} for the two eccentricity values that have been used in previous figures, i.e. $e=0.2$ and $e=0.4$. Let's describe first the results pertinent to $e=0.2$. In this case, the eccentricity is too small for the ionization to extend to the surface of the companion, and thus the effect on the stellar wind is limited (i.e. the value of the function $S$ in Eq.~\ref{eq:s} is close to unity). During the orbital phases when the NS is moving away from periastron toward apoastron (black lines), the relative velocity has a minimum around phase 0.1 and thus here we record the higher accretion luminosity. The relative velocity on the orbital phases when the NS is moving from apoastron toward the periastron (red lines) are characterized by relative velocities generally larger than those of the other phases, and thus the accretion luminosity along red lines is lower than that  during the black lines. When we consider the case of a larger eccentricity, the ionization has a substantially stronger effect and the accretion luminosity is significantly affected by the function $S$ (see Eq.~\ref{eq:s}). The luminosity is generally lower during most of the orbital phases compared to that obtained in the case $e=0.2$. This is because, although the wind velocity is low, the strong photoionization causes that only part of the wind material has a sufficiently high velocity to make it to the NS without falling back onto the supergiant. If we consider the orbital phases where the NS moves from periastron toward apoastron (black dashed line), we can see that the luminosity is remarkably low up to phase $\sim$0.55. At this phase, the photoionization is low because the NS is moving away from the companion, and the $S$ function approaches unity. The resulting mass accretion rate suddenly increases and then starts decreasing again in an exponential fashion given the fact that the NS is now far from the companion and the density of the symmetric wind is low. A similar situation occurs when the NS is moving from apoastron to periastron (red dashed line). In this case, the peak related to the S function going from unity to  lower values occurs closer to the supergiant compared to the case of the black dashed line. This is because of the asymmetric relative velocities in the phases crossed by the red and black lines.  

We thus conclude that when the photoionization of the wind by the X-rays from the accreting NS is strong enough, we can expect two asymmetric peaks in the X-ray luminosity during the orbital revolution at orbital phases intermediate between apoastron and periastron. The position and intensity of these peaks depend from all system parameters and in particular from the value of the critical ionization parameter $\xi_{\rm cri}$. For smaller values of $\xi_{\rm cri}$, the two peaks move toward the apoastron because the NS is able to strongly photoionize the stellar wind at lower X-ray luminosities. The opposite occurs for higher values of $\xi_{\rm cri}$, and we can see that already at $\xi_{\rm cri}$=300~erg~cm~s$^{-1}$ the two peaks get close to periastron (see Fig.~\ref{fig:lxacc}). The interesting aspect is that in a system with a smooth and symmetric stellar wind, the higher X-ray luminosity is never expected to occur precisely at periastron, but always slightly before and after. This is because the relative velocity between the stellar wind and the NS reaches a first minimum before the compact object is at periastron and a second even lower minimum slightly after the NS has left periastron. 

In the current calculations we cannot estimate the source variability on the short time scale (hundreds thousands of seconds) as we did not include in our simplified treatment the presence of clumps in the stellar wind or hydrodynamic instabilities of the accretion flow (see Sect.~\ref{sec:intro}). Therefore, the two asymmetric luminosity peaks have to be considered as structures in the source lightcurve profile that persist over many orbital revolutions (being an intrinsic effect of the eccentric orbit). Observationally, they should emerge not only while observing a single revolution of the source but also while folding together data from many revolutions (averaging out the short term variability).

\subsection{Applications to a few classical SgXBs and SFXTs}
\label{sec:applications}

The predictions about the variability of the luminosity along the orbit of an accreting NS in a SgXB have been described and summarized in Sect.~\ref{sec:lxacc}. Looking for observational evidences of peaks in the X-ray luminosity at specific orbital phases is not an easy task because for most of the SgXBs we are currently lacking orbital monitorings with sensitive instruments and spanning a sufficiently wide range of orbital cycles among many revolutions to average out the short term variability related to the clumpiness of the stellar wind. Furthermore, there are not (yet) many known SgXBs with a significant eccentricity and for most known systems we are lacking detailed measurements of the wind properties. For SgXBs the wind properties are generally inferred through indirect methods, such that different models of the supergiant wind (as the $\beta$-law introduced in Sect.~\ref{sec:wind} and the corresponding free parameters) are applied only to provide indicative values of the wind characteristics rather than obtaining reliable evaluations or even compare theoretical predictions with observational results. This limitation is due to the fact that the bulk of the stellar wind diagnostics are in the UV domain and the SgXBs are collectively characterized by relatively large columns of local absorbing material which make any measurement in that energy domain extremely challenging \citep[see, e.g., the discussion in][and references therein]{hainich20}. As the stellar wind properties are among the free parameters in our calculations, we highlight below the fact that in most cases we are able to carry out a qualitative, rather than a quantitative, comparison between the derived results in this paper and the observational properties available so far from SgXBs. The comparisons are in any case promising because they reveal some physical effects not considered before and thus deserve further developments along the lines that we later summarize in Sect.~\ref{sec:conclusions}.    
\begin{figure*}
\centering
\includegraphics[scale=0.35]{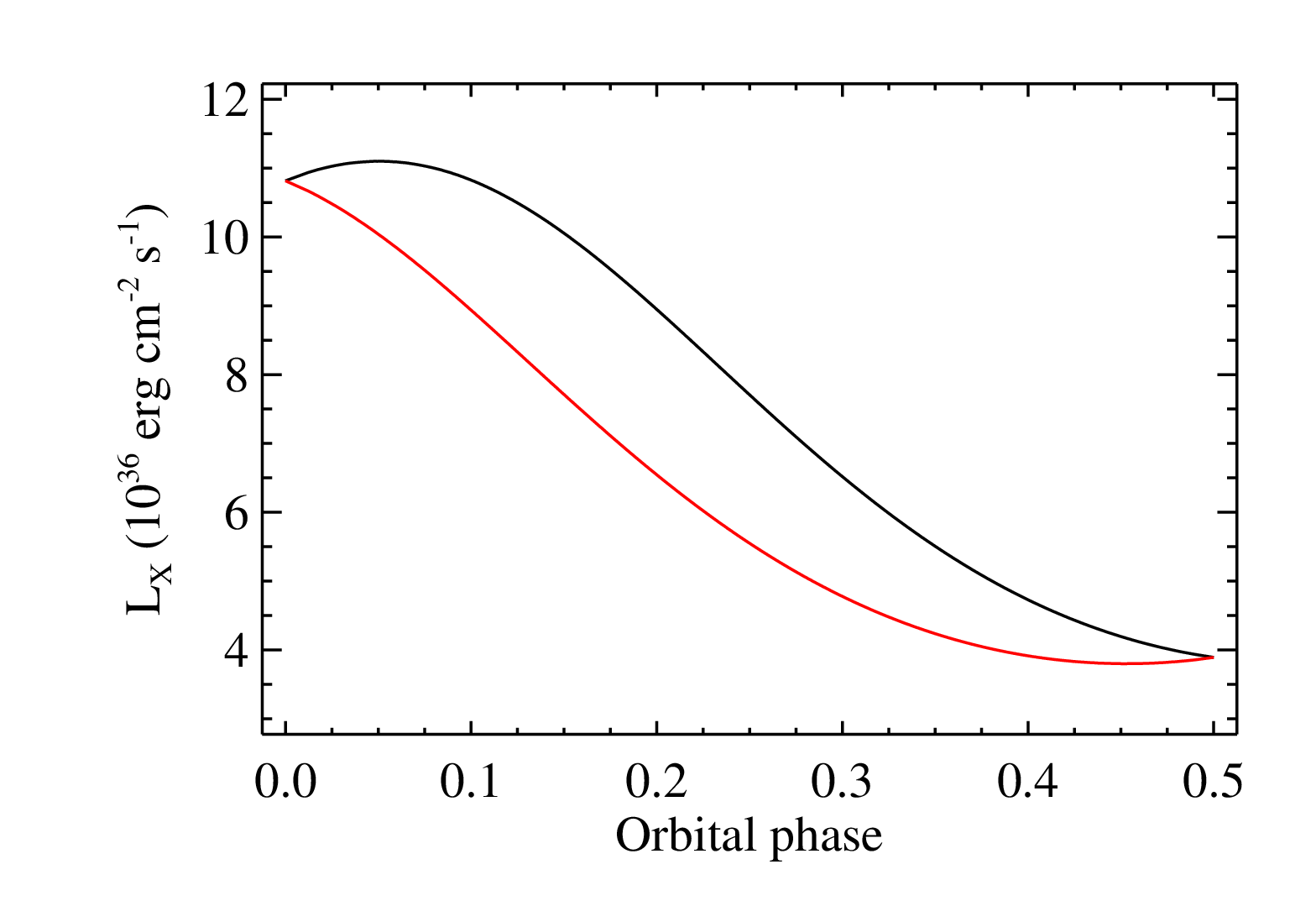}
\includegraphics[scale=0.35]{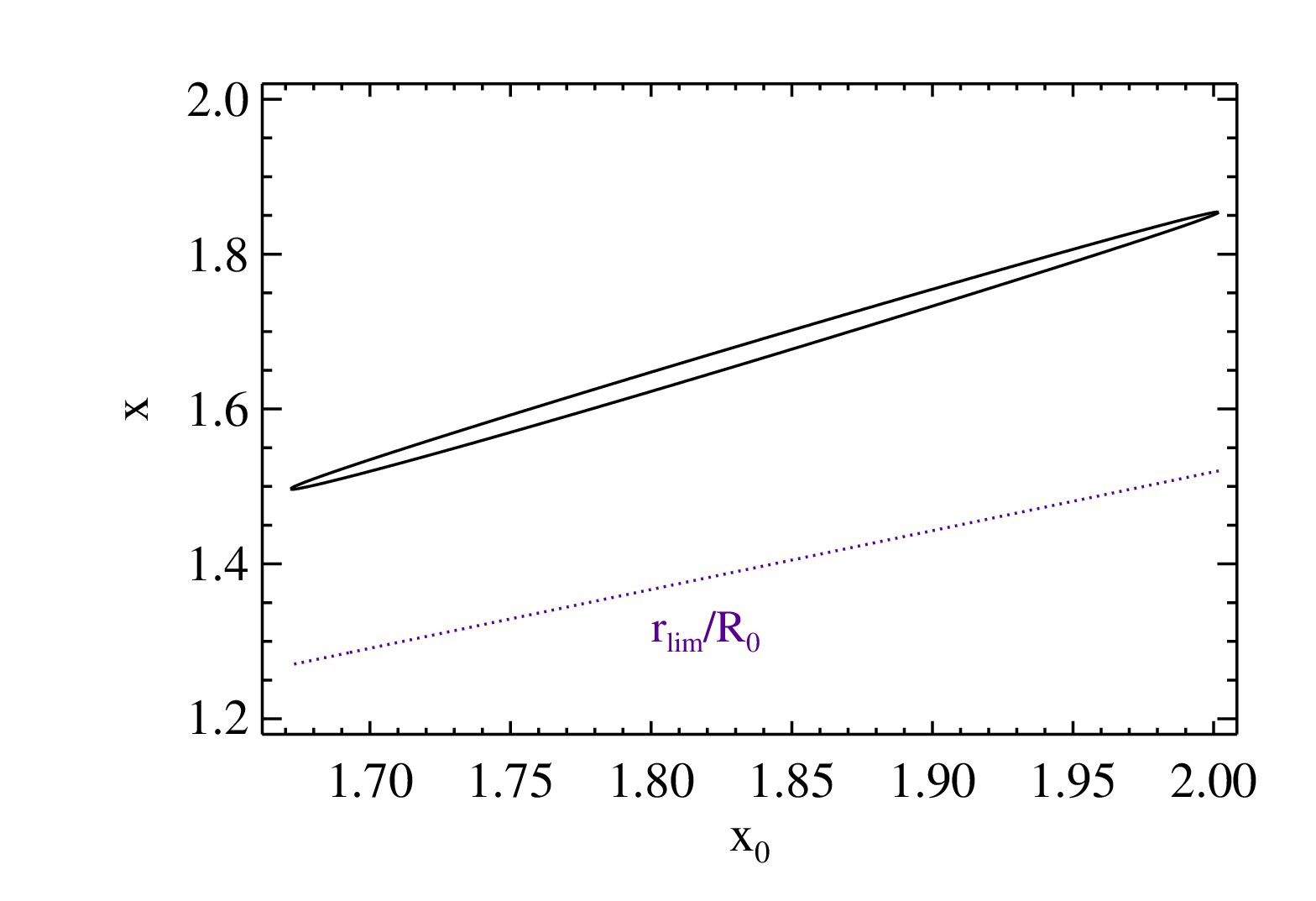}
\caption{{\it Left}: X-ray luminosity computed along the orbit of Vela~X-1 assuming the system parameters given by \citet{gimenez16}. The red line corresponds to the case where the NS is moving from apoastron to periastron (lower luminosities), while the black line corresponds to the inverse path (formally going from phase 0.5 to 1.0 although this is not changed for clarity on the x-axis). The luminosity along the red line is generally 20\% lower than that along the black line, compatible with the observations (see text for more details). {\it Right}: Position of the ionization radius as a function of the orbital separation calculated in the case of Vela~X-1.}  
\label{fig:velax1} 
\end{figure*}

\subsubsection{The classical SgXB prototype Vela\,X-1}

We consider first the case of the classical SgXB prototype, Vela~X-1. This is a test-bench case to verify the application of our calculations to the best known system among all SgXBs. In Vela\,X-1, a NS is orbiting with a period of 8.9~d a B supergiant and the measured eccentricity of the NS orbit is of 0.0898 \citep[see, e.g.][and references therein]{walter2015, falanga15}. Large uncertainties exist concerning the properties of the B supergiant and its wind, especially due to the fact that these parameters are affected by the X-ray photoionization and thus their derivations are somewhat model-dependent. For our application, we consider here the parameters reported by \citet[][see their Table~1]{gimenez16}. Although these parameters were slightly refined in a more recent publication \citep{sander18}, these latter authors opted for a more advanced approach for the definition of the stellar wind structuring rather than adopting the simpler  $\beta$-law approximation and thus their approach cannot be easily reconciled with our calculations (see Sect.~\ref{sec:wind}). Although the treatments in \citet{gimenez16} and \citet{sander18} are substantially different, some of the key properties of the stellar wind that they obtain (and that are of interest for our calculations, e.g. the mass loss rate) are compatible to within the uncertainties of the derivations. As the B supergiant mass, radius, mass loss rate, wind terminal velocity, as well as the orbital parameters are all fixed, the main free parameter in our calculations to reproduce the observed X-ray luminosity from Vela\,X-1 remains $\xi_{\rm cri}$. 

In order to achieve a good match between the X-ray luminosity computed in our calculations and that observationally measured at different orbital phases\footnote{Further fine tuning of the X-ray luminosity to match more accurately the observations is possible but it is beyond the scope of the current paper given the simplifications adopted in our calculations. These includes the assumption through the Eq.~\ref{eq:lacc} of a 100\% efficiency in converting all energy of the accreted matter into X-ray luminosity.}, we had to assume fairly large values of the critical ionization parameter, i.e. $\xi_{\rm cri}$$\simeq$10$^4$. The results obtained in this case are reported in Fig.~\ref{fig:velax1}. The plot of the X-ray luminosity along the orbit shows a difference of about 15\% when the NS is approaching (red line, lower luminosity) and leaving (black line, higher luminosity) periastron. An asymmetry of the luminosity around periastron at a qualitatively similar level can be appreciated in several published plots of the long-term averaged emission from the system as a function of the orbital phase, as those reported in Fig.~1 from \citet[][hereafter NZ14]{nunez14} and Fig.~3 from \citet{kre19}. These plots also show, in agreement with our results in the left side of Fig.~\ref{fig:velax1}, that the luminosity in the orbital phases when the source is going from apoastron to periastron (phases 0.7 to 0.2 in NZ14) is generally about 20\% lower than in the orbital phases when the source moves away from periastron and goes back toward apoastron (phases 0.2 to 0.7 in NZ14). Note that a more precise comparison between the estimated and observed \% of asymmetry is difficult to be carried out as in Vela~X-1 the approach toward periastron occurs close to the egress from the X-ray eclipse and there are no sufficiently simultaneous high sensitivity and long-term monitoring observations of this source to firmly establish the exact orbital phase at which the egress from the eclipse is complete and the source comes back to its X-ray flux outside occultation (given also the short term X-ray variability affecting short pointed  observations outside the eclipse). The choice of a relatively high $\xi_{\rm cri}$ in our calculations implies that the radius of the Str$\ddot{\rm o}$mgren sphere in Vela\,X-1 does not extend too close to the supergiant. This is shown in the right panel of Fig.~\ref{fig:velax1}, together with the value of $r_{\rm lim}$. The latter is smaller than $x$ at all orbital phases, meaning that the photoionization of the wind remains mild. This is compatible with the more detailed findings from \citet{sander18} who found that the largest slow down of the B supergiant wind in Vela\,X-1 occurs close to the NS location rather than to the supergiant surface. 

Vela~X-1 is also one of the sources among the SgXBs displaying frequent off-states. These events in Vela~X-1 have been clearly detected throughout the long-term monitoring observations performed with the \inte\ satellite \citep{kreykenbohm08} and found to last several hundreds of seconds up to a ks. The drop in luminosity observed by \inte\ during the off-states is typically of a factor of $\sim$10, with the source being undetected during the occurrences of these events. When observed with higher sensitivity instruments, as the XIS on-board \suzaku,\ the decrease in luminosity during the off-states is revealed to be around a factor of $\sim$20, with the residual X-ray emission still being modulated on the pulse period of the NS \citep{doroshenko10}. This means that some limited accretion is still taking place during the off-states. A statistical study of the distribution of the off-states as a function of the orbital phase in Vela~X-1 was presented by \citet{sidoli14}. These authors found that the off-states are occurring at all orbital phases, with a concentration of events during the ingresses and egresses from the eclipses. However, at these specific orbital phases the off-states detected by \inte\ are most likely related to the effect of the accretion wake and thus have a different origin compared to those at all other orbital phases. 

In the simplified calculations presented in the previous sections, the off-states are suggested to be produced when the X-ray emission of the NS is sufficiently high to temporarily stop the stellar wind and inhibit accretion. Since our calculations showed that a high value of $\xi_{\rm cri}$ is needed in Vela\,X-1 to explain the long-term averaged X-ray luminosity at different orbital phases, the remaining possibility to explain the off-states within the photoionization model is to assume that the photoionization is not triggered by  the enhanced accretion close to periastron but rather it occurs at any orbital phase during the brighter emission episodes associated with the accretion of clumps. In our presently simplified approach the wind is assumed smooth and symmetric, and thus we cannot quantitatively test this hypothesis but it is well known that the stellar wind around the NS in Vela~X-1 is characterized by massive clumps \citep[see, e.g.,][and references therein]{fuerst10} and thus this is definitively a test of the photoionization scenario to be verified in future versions of our calculations. The results reported in Fig.~1 from \citet{sidoli14} suggest that most of the off-states away from the ingresses and egresses from the eclipse occur in coincidence with the brightest emission episodes. This would thus support our proposed scenario, although a statistical study of the correlation between the occurrence of an off-state and the X-ray luminosity of Vela\,X-1 would still need to be carried out\footnote{We note that a similar role to that of the clumps to initiate a substantial photoinoization could be played by the passage of the neutron star through a region where the wind along the line of sight to the massive companion is endowed with a particular low density/fast velocity. This could also facilitate the photoionization to larger extends and result in off-states. The discrimination between the clump and the low density/fast velocity region case could be tested by means of accretion column density measurements before, during, and after the off-state, although in most cases the statistics of the data does not allow us to discriminate between different intertwined absorption components along our line of sight to the source.}. The presence of pulsations during the off-states is also not in contradiction with the idea that these states are associated to episodes of strong photoionization. Indeed, while the photoionization blocks the stellar wind for a time comparable with the local crossing time between the supergiant and the NS, the latter is still able to accrete from the surrounding wind material that has not been cleared out yet. The clearing out is also expected to take place within a similar time scale and thus a strong reduction of the accretion rather than a complete turn off of the X-ray emission seems plausible within the photoionization model for the off-states. 

We conclude that the application of our calculations to the case of Vela~X-1 is qualitatively compatible with the observations and we are able to provide an interpretation for the long-term asymmetry of the system luminosity while approaching and moving away from periastron. We are also able to provide a suitable explanation for the off-states with the photoionization model. A more quantitative comparison is needed during the future implementation of the missing ingredients in our calculations (e.g., the clumpy wind).  
\begin{figure*}
\centering
\includegraphics[scale=0.35]{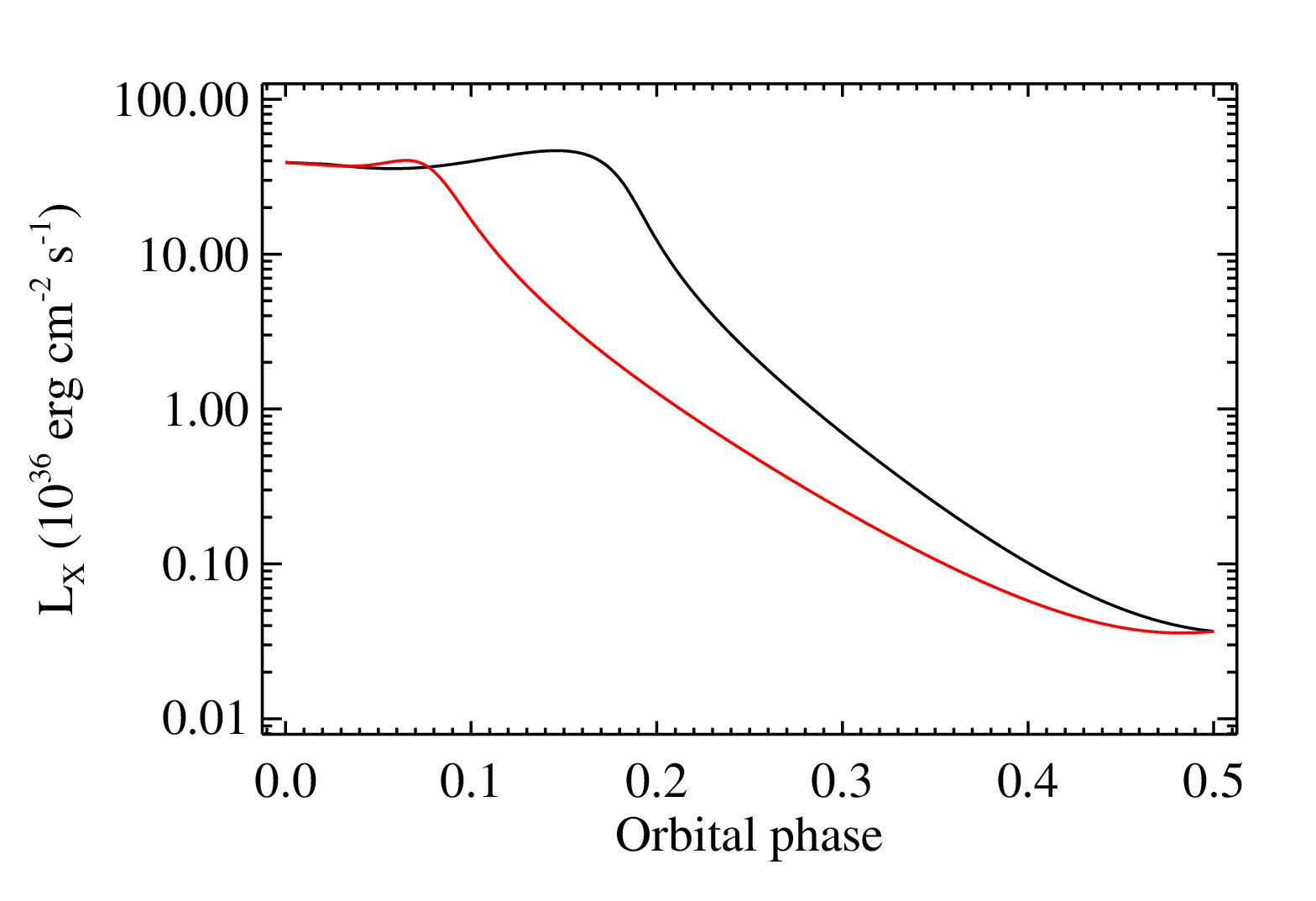}
\includegraphics[scale=0.35]{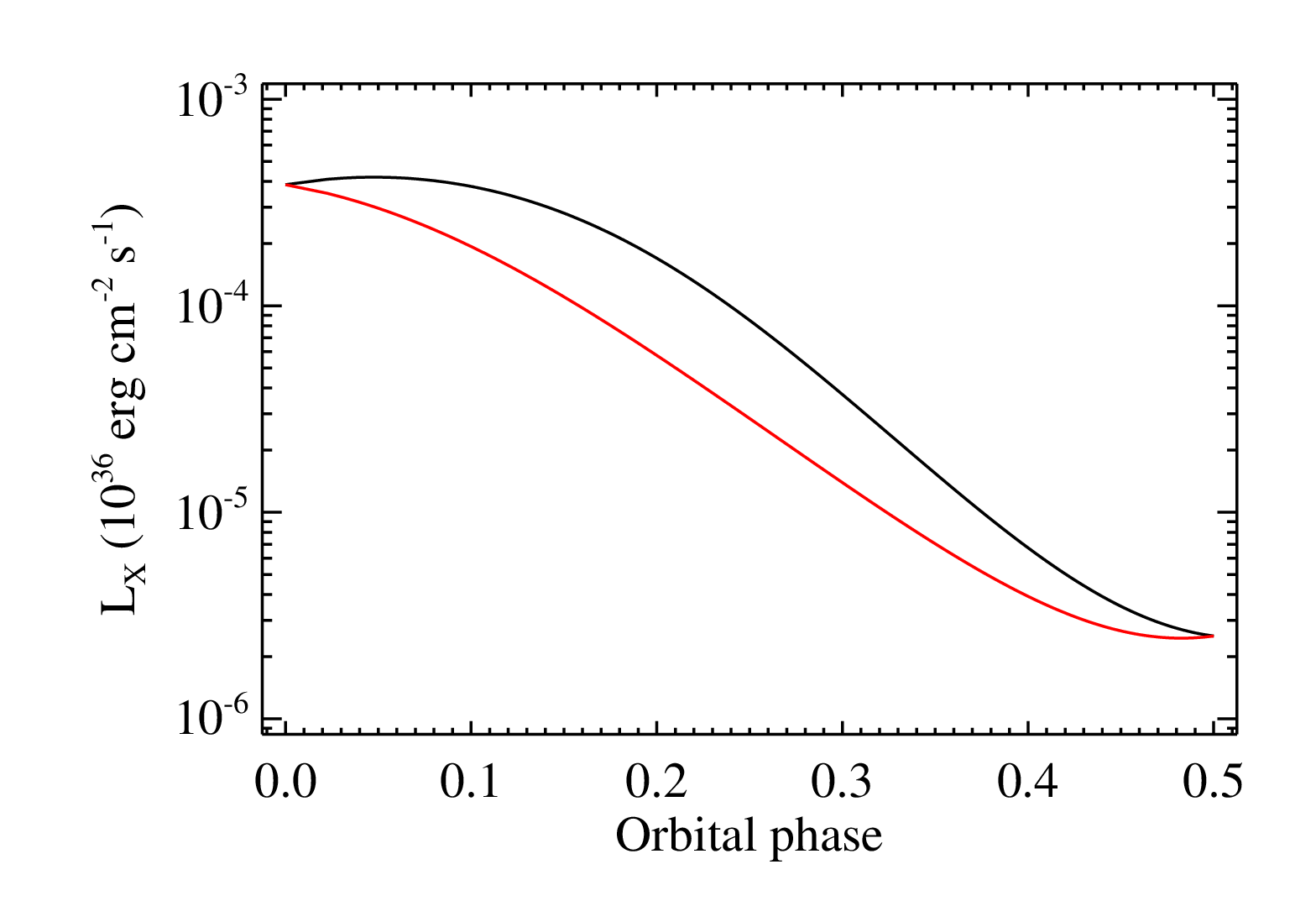}
\caption{{\it Left}: X-ray luminosity computed along the orbit of the SFXT IGR\,J08408-4503 assuming the system parameters discussed in the text. As in the other figures in this paper, the red line corresponds to the case where the NS is moving from apoastron to periastron (lower luminosities), while the black line corresponds to the inverse path. {\it Right}: the same as the left panel but including the factor $\chi=7\times10^{-5}$ in Eq.~\ref{eq:lacc} to mimic the effect of the inhibition of accretion and reproduce the average luminosity measured around periastron.}   
\label{fig:J08408} 
\end{figure*}

\subsubsection{SgXBs with pronounced eccentricities: 4U\,1907+09 and GX\,301-2}

Although Vela\,X-1 is an important test-bench, the application of our calculations to SgXBs with more eccentric orbit could in principle help us better in verifying the predicted level of asymmetry in the X-ray luminosity while approaching and exiting periastron (the level of asymmetry is a strong function of the eccentricity, given the dependence of $v_{\rm rel}$ on this parameter, see Fig.~\ref{fig:coord}). Among the wind-accreting classical SgXBs, high eccentricities have been firmly measured in both 4U\,1907+09 \citep[$e=0.28$][and references therein]{doroshenko12} and GX\,301-2 \citep[$e=0.46$][and references therein]{doroshenko10}. The problem of the application of our calculations to these sources is that the former is known to display a bright regular flare right at periastron \citep[see, e.g.,][and references therein]{sainer12} and the latter undergoes similar flares just before entering periastron \citep[see, e.g.,][and references therein]{fuerst11}. Both sources lack sufficiently deep and long observations around periastron with sensitive X-ray instruments to study in detail any possible asymmetry in the emitted luminosity while entering or recessing periastron, but there are important considerations that can be made from our calculations for both sources concerning their peculiar activity close to the periastron. As shown in our Fig.~\ref{fig:coord}, a purely wind accreting system characterized by an eccentric orbit is expected to have on the long term (i.e. when the rapid variability related to clumps has been averaged out) a lower luminosity while approaching the periastron compared to the receding phase. This is because the relative velocity $v_{\rm rel}$ is higher during the approach to periastron and lower during the recess. This means that a flare before or during periastron is not expected unless the system switches to a different type of accretion, e.g. due to the formation of a temporary accretion disk. As discussed in Sect.~\ref{sec:disk2}, a temporary accretion disk is more likely to form when the separation between the NS and the companion is close to its minimum and/or where the orbital and stellar wind velocity nearly cancel each other. In the case of GX\,301-2, the flare occurs before the periastron when neither the distance between the two stars nor $v_{\rm rel}$ is at minimum. Based on our calculations, we thus consider it unlikely that GX\,301-2 switches from wind to disk accretion during the pre-periastron flare. Alternative models, where the pre-periastron flare is due, e.g., to the NS overtaking its normally preceding accretion stream seem thus favored \citep{leahy08}. The periastron flare displayed by 4U\,1907+09 could be instead likely related to the formation of a temporary accretion disk, in agreement with previous suggestions in the literature  \citep{zand98,sainer12}. 

Interestingly, both 4U\,1907+09 and GX\,301-2 are also known to display off-states. In the case of 4U\,1907+09, \citet{doroshenko12} reported the analysis of several off-states during a high sensitivity pointed observation with \suzaku\ and showed that the source emission properties during these states are very similar to those observed in Vela~X-1, including the drop in luminosity compared to the average emission, the spectral variation during and outside the off-states, as well as the detection of pulsations during these event \citep[note that previous reports on the analysis of the off-states from 4U\,1907+09 did not succeed in revealing pulsations due to the limited sensitivity of the employed instruments; see the discussion in][]{doroshenko12}. The duration of the off-states in 4U\,1907+09 was also measured to be of the order of several hundreds of seconds, as for Vela\,X-1. Given all these similarities between the off-states in 4U\,1907+09 and Vela~X-1, the same considerations as in the above two paragraphs also apply to the former source. As further reported by \citet{doroshenko12}, the probability of finding 4U\,1907+09 in an off-state is lower when the source is close to periastron.  This is not in contradiction with the wind ionization model because, although the model  predicts a higher concentrations of off-states when the X-ray luminosity from the source is higher, the system is likely to switch to the disk accretion mode around periastron and thus it is expected that the ionization of the wind does not cause a rapid drop of the accretion rate (which is mostly regulated by viscosity processes within the temporary disk). 

In the case of GX\,301-2, the situation is much less clear as only two detections of  off-states have been reported in the literature. According to the results published by \citet{gogus11}, the first off-state detected from the GX\,301-2 displays similar properties to those from the previously reported sources, as the luminosity underwent a drop by a factor of $\sim$10 and the duration is of about 1~ks. The event occurred at phase $\sim$0.7, when the source was half-way from apoastron to the periastron passage. At odds with the off-states from the other sources, pulsations seem to disappear in the case of GX\,301-2, at least for a single pulse period. These results were confirmed also for the second off-state detected from the source and reported by \citet{fuerst11}. These authors suggested that the off-states in GX\,301-2 are likely due to a substantial drop in the accretion rate, rather than an obscuration event due to the lack of any evidence of an increased absorption column density during these events. This conclusion would thus well match the expectations of the wind ionization scenario, although higher resolution X-ray observations could help investigating the presence of strongly ionized material close to the NS. 

\subsubsection{IGR\,J08408-4503: the only SFXT with a firmly confirmed eccentric orbit}

We conclude Sect.~\ref{sec:applications} with the application to the SFXT IGR\,J08408-4503. This is the only SFXT for which an accurate determination of the orbital characteristics, including a high eccentricity, has been securely confirmed. This makes IGR\,J08408-4503 an interesting source for the application of our calculations presented above. By exploiting optical observations, \citet{gamen15} has measured for this SFXT an orbital period of 9.5436$\pm$0.0002~d and an eccentricity of 0.63$\pm$0.03. The massive star in this system is identified as a O8.5 Ib-II(f)p supergiant at a distance of 2.21$^{+0.14}_{-0.16}$~kpc \citep{bailer18}. Similarly to all other SFXTs, IGR\,J08408-4503 spends most of its lifetime in a quiescent state \citep[about 67\%;][]{romano14} at a luminosity of 10$^{32}$-10$^{33}$~erg~s$^{-1}$ 
and displays only sporadically outbursts reaching luminosities of a few 10$^{36}$~erg~s$^{-1}$ \citep[see, e.g.,][and references therein]{bozzo10,bodaghee11,bozzo15}. Although the system is known to preferentially emit outbursts close to periastron (orbital phase 0), these events do not occur regularly and it happened that long lasting observations around this specific orbital phase found the source at a quiescent luminosity level \citep[$\sim$4$\times$10$^{32}$~erg~s$^{-1}$;][]{sidoli10}. A complete study of the distribution of all outbursts observed from the source as a function of the orbital phase was reported by \citet{ducci19}. These authors showed that all outbursts occur when the compact object, most likely a NS, moves away from periastron between phases 0-0.15 and when it approaches again the periastron between phases 0.85-1.0. Based on the companion star physical properties published by \citet{gamen15}, \citet{ducci19} suggested that in this source accretion might occur through a disk due to the onset of the Roche-Lobe overflow. The physical properties of the supergiant  in IGR J08408-4503 reported in \citet{hainich20} support the RLO at periastron scenario only within a confidence level of 90\% (see their Table 4 and their Section 8), thus leaving room to other hypotheses. 

In order to apply our calculations to this source we assume the higher supergiant mass of 13~$M_{\odot}$ within the uncertainty range presented by \citet{hainich20} that allows the massive star to reside within its Roche-Lobe radius for any orbital phase \citep[we estimated the Roche-Lobe radius as indicated by][]{ducci19}. This ensures us that accretion always occurs through the stellar wind and not through an accretion disk. The other stellar parameters, including the mass loss rate, are fixed to the best determined values in \citet{hainich20}. We show in Fig.~\ref{fig:J08408} (left) the result of our calculations assuming the standard parameters for a NS (a mass of 1.4~$M_{\odot}$ and a radius of 10~km) and a value of $\xi_{\rm cri}$=1.3$\times$10$^3$. The latter parameter has been set in order to have the maximum luminosity of the system in the orbital phase interval $\sim$0.85-1.15 where most of the outbursts are observed. Around this orbital phase interval, according to our calculations, the luminosity is enhanced by the photoioization of the stellar wind which is decreasing the wind velocity and increasing the accretion rate onto the NS. The figure shows that if IGR\,J08408-4503 was to behave as a classical wind-fed SgXB, its average X-ray luminosity all along the orbit (and especially around periastron) should be much larger than what is actually observed. The so-called ``sub-luminosity'' issue is a well known property of all SFXTs \citep[see, e.g.,][]{bozzo15} and the most accredited  scenario to explain this behavior is to assume that some mechanism is at work in these objects to inhibit the accretion for most of the time (see Sect.~\ref{sec:intro}). Although in the present version of our calculations we did not yet include a complete treatment of the magnetospheric interactions that could led to such inhibition, we can mimic the effect of the inhibition of accretion by adding a factor $\chi\ll1$ in front of $\dot{M}_{\rm acc}$ in Eq.~\ref{eq:lacc}. In order to recover the observed luminosity of 4$\times$10$^{32}$~erg~s$^{-1}$ around periastron \citep{sidoli10}, a value of $\chi\simeq7\times10^{-5}$ is required. The outcome of this calculation is shown in the right side of Fig.~\ref{fig:J08408}. Here the profile of the luminosity variation is much smoother than in the previous case as the low average X-ray luminosity (due to the inhibition of accretion) does not cause a significant ionization of the stellar wind. The relatively large variability of the expected luminosity along the orbit is only related to the changing distance between the compact object and the supergiant star. This is, however, much lower than the dynamic range of the X-ray luminosity measured in IGR\,J08408-4503 and in all other confirmed SFXTs (10$^5$-10$^6$). 

Our conclusion is thus that the mechanism(s) inhibiting accretion in IGR\,J08408-4503 (and for similarity in likely all other SFXTs) shall be very effective in screening the NS from the mass inflow rate coming from the supergiant. Among the different models presented in the literature, the magnetic gating discussed by \citet{bozzo08} is the most efficient mechanism blocking accretion. This would require simultaneously a relatively slow spinning NS ($\gtrsim$1000~s) and a very strong magnetic field ($\sim$10$^{14}$~G). This possibility cannot be excluded at present, because the spin period of the NS hosted in IGR\,J08408-4514 and its magnetic field intensity are not known. This condition is similar to all most extreme SFXTs as the only confirmed spin period is the one of the peculiar SFXT IGR\,J11215-5952 showing strictly periodic outburst and the only evidence for a cyclotron line in these objects was reported for the SFXT IGR\,J17544-2619 but not confirmed by subsequent observations \citep{bhalerao15,bozzo17}.

\section{Conclusions}
\label{sec:conclusions}

In this paper we reported on the first step toward a more-comprehensive semi-analytical approach to study the process of wind-accretion onto a neutron star in supergiant X-ray binaries with eccentric orbits. In this first version of the calculations, we applied several simplifications, including the assumption of a spherically symmetric wind for the supergiant, the adoption of a simplified accretion scenario with no magnetospheric interactions, and we neglected the effect of the Corioli's force. Compared to previous semi-analytical approaches published in the literature, we improved by including  the effect of the relative velocity between the orbiting neutron star and the stellar wind, as well as a self-consistent treatment of the X-ray photoionization of the stellar wind and the presence of a non-negligible orbital eccentricity. Compared to the relevant numerical simulations made available so far, the advantage of our calculations is to provide a relatively simple treatment that allows us to identify key aspects of the effect of eccentric orbits on the accretion process in NS SgXBs never studied before (and not yet included in numerical simulations). The equations we provide allow the reader to identify the physical origin of some interesting effects that can be (qualitatively) compared with the observations of NS SgXBs in the X-ray domain. Indeed, while further improvements and extensions of our calculations are planned for forthcoming publications, we could already obtain interesting results while computing the expected X-ray luminosity of a NS SgXB as a function of the orbital phase. These findings could inspire future additional numerical efforts devoted to deepening our understanding of the wind-fed systems.  

We showed first in Sect.~\ref{appendix:ho87} that the so-called high and low luminosity solutions reported by previous calculations in the literature \citep{ho87,krtika12,karino14} about the wind photoionization effects in SgXBs are unlikely to coexist simultaneously. A system endowed with a sufficiently high eccentricity can rather experience both of them when the neutron star approaches or moves away from the periastron. This insight provides the missing physical motivation that allows us to understand the conditions under which either the low or the high luminosity solution applies. 

Convolving the effect of the X-ray photoionization with the modulation expected from a SgXBs with an eccentric orbit provided clear predictions for the long-term average  luminosities along the revolution (once the short term variability due to the clumpiness of the wind is averaged out by integrating over, e.g., many orbital periods). We identified that in wind-fed systems with eccentric orbits two peaks in the X-ray luminosity are expected slightly before and after the periastron passage. The precise orbital phase of the peaks largely depends on the assumed critical ionization parameter. For lower values of $\xi_{\rm cri}$, the peaks move toward the apoastron because the NS is able to efficiently photoionize the wind already at lower X-ray luminosities. The opposite occurs for higher values of $\xi_{\rm cri}$. There is also an expected asymmetry of the X-ray luminosity displayed by the system experiencing only a negligible to mild photoionization of the wind before entering and after exiting periastron. This is due to the difference in $v_{\rm rel}$ when the NS approaches its companion or moves away from it. We compared these predictions with observational findings on the known classical SgXBs endowed with a non-negligible eccentricities, i.e. Vela\,X-1, 4U\,1907+09, and GX\,301-2, and with the only SFXT for which an orbital eccentricity has been firmly measured, IGR\,J08408-4503.

In the case of Vela\,X-1, the measured eccentricity is relatively small and thus we found for this source that the photoionization of the wind is mild at all orbital phases. No photoionization-induced peaks of the X-ray luminosity are expected. Our predictions about the luminosity asymmetry while entering and existing periastron is quantitatively in agreements with observations of Vela\,X-1. For 4U\,1907+09 and GX\,301-2, the search for the peaks in the X-ray luminosity and/or the luminosity asymmetry is made less straightforward by the peculiar X-ray flaring behavior of these sources close to periastron. We could conclude, thanks to our calculations, that only for 4U\,1907+09 the flare at periastron is likely to result from the formation of a temporary accretion disk around the compact object, while for GX\,301-2 alternative models are preferred. In the case of IGR\,J08408-4503, our calculations remarked clearly that efficient mechanisms to inhibit the accretion are required to reproduce the source observed X-ray luminosity, especially at periastron. Even the modulation induced by the large eccentricity of this system is far from being comparable (by orders of magnitudes) to that usually displayed by all SFXTs. The fact that the ensemble of these sources is generally much sub-luminous compared to classical SgXBs is well known and in future versions of our calculations we aim at testing the effect of the inhibition of accretion in a more quantitative way by including an extended treatment of magnetospheric interactions.  

We proposed that the strongest wind photoionization events could give rise to the so-called off-states, displayed by many SgXBs. During these states there is a drastic reduction of the accretion rate for a period that could be as long as a few ks. In our interpretation of these events, the wind photoionization causes a drop of the main (and sole) acceleration mechanism of the wind and thus a major reduction of the mass inflow rate toward the NS. When the photoionization extends up to the photosphere of the supergiant, the mass inflow rate toward the NS is interrupted and X-rays are drastically reduced. The NS can still undergo some residual accretion from the material floating around that has not been cleared up yet, but the low resulting X-ray luminosity is not sufficient to pursue the photoionization. Therefore, after a timescale roughly comparable with the wind travel time between the NS and the supergiant, the wind is expected to be restored and accretion to begin again. As the travel time is usually of a few hundreds of seconds, this model for the off-states provide a qualitatively good match with the observations of several SgXBs displaying off-states (taking into account also all simplifications to the calculations presented here). This does not exclude that other proposed mechanisms for the production of the off-states, as, e.g., effects associated to the hydrodynamics of the stellar wind or the presence of magnetic/centrifugal gates associated to the NS rotation and magnetic field \citep{manousakis15, doroshenko10} could be at the origin of these events but provide an additional plausible explanation to be tested further by future observations.

\section*{Data availability}

The data underlying this article are available in the article.

\section*{acknowledgements}
We thank an anonymous referee for the constructive comments that helped us improving the paper.

\bibliography{accretion.bib}{}
\bibliographystyle{mnras}

\end{document}